\documentclass[11pt]{article}
\usepackage[margin=.8in]{geometry}                   
\usepackage{graphicx}
\usepackage{subcaption}
\usepackage{amssymb}
\usepackage{amsmath}
\usepackage{epstopdf}
\usepackage{enumerate}
\usepackage{cancel}
\usepackage{authblk}
\usepackage[numbers]{natbib}
\usepackage{xcolor}

\title{The universality of skipping behaviours on music streaming platforms}

\author{Jonathan Donier\footnote{jdonier@spotify.com}}
\affil{Spotify}

\begin{document}
\maketitle

\begin{abstract}
A recent study of skipping behaviour on music streaming platforms has shown that the skip profile for a given song -- i.e. the measure of the skipping rate as a function of the time in the song -- can be seen as some intrinsic characteristic of the song, in the sense that it is both very specific and highly stable over time and geographical regions. In this paper, we take this analysis one step further by introducing a simple model of skip behaviours, in which the skip profile for a given song is viewed as the response to a small number of events that happen within it. In particular, it allows us to identify accurately the timing of the events that trigger skip responses, as well as the fraction of users who skip following each these events. Strikingly, the responses triggered by individual events appears to follow a temporal profile that is consistent across songs, genres, devices and listening contexts, suggesting that people react to musical surprises in a universal way.
\end{abstract}

\section{Introduction}

At the heart of music creation lies the interaction between an artist and its public. This interactivity is lost with artefacts such as vinyls or CDs, which turn listening into a rather passive activity, and break the feedback from fans to artists. The advent of the digital era, which culminates today in the ubiquity of music streaming, is opening again the possibility of a feedback cycle by providing intuitive ways for listeners to interact with a song and the ability for the platforms to record such interactions. Indeed, proactive listening behaviours (skipping, scrubbing, ...) are becoming so prominent \cite{lamerethedrop, lameretheskip} -- and, as we shall see, predictable -- that their analysis promises to reveal some deep insights about the music, which could in turn fuel the music making process.

A recent study \cite{montecchio2019skipping} has shown that the skip profile of a song -- i.e. the measure of the skipping rate as a function of the position in the song -- is both very specific to the song and highly stable across time and geographical regions, as if it were some intrinsic property of the music. More generally, it was repeatedly observed \cite{lameretheskip, montecchio2019skipping} that skip profiles exhibit a universal U-shape pattern with three phases:

\begin{enumerate}[(i)]
\item a high skip rate at the beginning of the song followed by a sharp power-law decay, as it takes up to a few seconds for people to decide whether or not they want to listen to the song,
\item a permanent regime with low skipping rate interspersed with spikes, whose temporal positions have been shown in \cite{montecchio2019skipping} to correspond to salient events such as musical transitions (the beginning of a chorus, the appearance of a voice or a new instrument, a variation in intensity...),
\item an increase in the skipping rate as the end of the song approaches, as users look forward to the next song.
\end{enumerate}

Motivated by these results and by the apparent universality of the above patterns across songs, geographical regions and time periods, we develop a model for skipping behaviour as a function of some (\textit{a priori} unknown) underlying events in the song. In particular, we find that the patterns constitutive of the three phases above (namely, the initial decay, the spikes in permanent regime, and the increase at the end) can be accurately modelled by some universal response functions (also called \emph{kernels}), that are only parametrized by their timings and amplitudes. Notably, we find that the distribution of response times to musical events decays slowly at large times, consistent with a number of behavioural studies that have shown that human response times on various tasks can be approximated by a log-normal distribution \cite{schnipke1999representing, buzsaki2014log, holden2009dispersion, holden2002fractal, van2003self, thissen1983timed}. Fitting this model to real data enables one to perform a quantitative and interpretable comparison of skip profiles, which could be further analyzed by artists who wish to dissect how their songs are being received by their public -- for example, by measuring the fraction of listeners that are lost over a particular musical transition. 

We start by introducing the model for skipping behaviour in Section \ref{sec:model} and show that the timings of the kernels indeed correspond to musical events -- typically transitions -- as first observed in \cite{montecchio2019skipping}. In Section \ref{sec:universal},  we test empirically the modelling hypothesis that the shape of the skip response triggered by musical events does not depend on the musical genre or the listening context. In Section \ref{sec:spikes}, we perform a day-by-day analysis of the model parameters, and show that they are a very stable and distinctive characteristic of a song, which again confirms the findings of \cite{montecchio2019skipping} as well as the relevance and stability of our model. A qualitative and quantitative analysis across various characteristics of the music (such as genre, stream count and listening context) is additionally performed in Appendix \ref{sec:analysis}.
\section{A model for skipping behaviour}\label{sec:model}

\subsection{Skip profiles as Poisson processes}

For a given song of length $T$ with a total of $N$ streams, let $N(t)$ be the number of sessions who were still active at time $t^-$, and $s(t)$ the associated skip profile such that the number of skips in the interval $[t_0; t_1]$ is equal to: 

\begin{equation}
N(t_1^+) - N(t_0^-) := \int_{t=t_0}^{t_1}s(t)\mathrm{d}t.
\end{equation}

\noindent We model the skip profile as a Poisson process with time-varying intensity $\lambda(t)$, such that the fraction of sessions that were active at time $t^-$ but became inactive (i.e. where the user skipped) between $t$ and $t + \mathrm{d}t$ is $\lambda(t)\mathrm{d}t$. Given a parametrization of the model space $\lambda_\theta(t)$, the optimal set of parameters $\theta^*$ can then be found by applying the following optimization program:

\begin{equation}
\theta^* := \underset{\theta}{\text{argmax}}~\mathrm{log}\left(\mathrm{d}\mathbb{P}\left[\lambda_\theta(t) \mid s(t) \right] \right),
\end{equation}

\noindent where the posterior log-probability of the intensity $\lambda(t)$ can be written as:

\begin{equation}
\begin{aligned}
\mathrm{log}\left(\mathrm{d}\mathbb{P}\left[\lambda(t) \mid s(t) \right] \right) &= \mathrm{log}\left(\mathrm{d}\mathbb{P}\left[ s(t) \mid \lambda(t) \right] \right) + \mathrm{log}\left(\mathrm{d}\mathbb{P}\left[ \lambda(t) \right] \right) - \mathrm{log}\left(\mathrm{d}\mathbb{P}\left[ s(t) \right] \right)\\
&\propto \underbrace{\mathrm{log}\left(\mathrm{d}\mathbb{P}\left[ s(t) \mid \lambda(t) \right] \right)}_{\text{log-likelihood}} +  \underbrace{\mathrm{log}\left(\mathrm{d}\mathbb{P}\left[ \lambda(t) \right] \right) }_{\text{model prior}}.
\end{aligned}
\end{equation}

\noindent The model prior, which acts as a regularization term, depends on the modelling hypotheses, which we will describe in Section \ref{sec:prior}. The log-likelihood term, on the other hand, can be derived directly from the Poisson hypothesis above.

\subsection{Log-likelihood}

 Let us assume that the skip decisions are independent\footnote{This is true for different people, probably less so for different streams by the same person.} and note $s_i(t)$ the skip profile corresponding to the $i$-th listening session (such that $s(t) = \sum_{i=1}^N s_i(t)$) and $\tau_i$ the time at which session $i$ stopped (with $\tau_i = T$ if the song was not skipped). Then, the probability of observing the skip profile $s(t)$ given an intensity $\lambda(t)$ is given by:

\begin{equation}
\begin{aligned}
\mathrm{d}\mathbb{P}\left[ s(t) \mid \lambda(t) \right] &= \prod_{i=1}^N \mathrm{d}\mathbb{P}\left[ s_i(t) \mid  \lambda(t) \right]\\
&= \prod_{i=1}^N \mathbb{P}\left[\text{no skip at } \tau_i \mid  \lambda(t) \right] \mathrm{d}\mathbb{P}\left[ \text{skip in } (\tau_i , \tau_i + \mathrm{d}\tau_i)  \mid \text{no skip at } \tau_i ,  \lambda(t) \right]\\
&= \prod_{i=1}^N e^{-\int_0^{\tau_i} \lambda(s)  \mathrm{d}s}  \lambda(\tau_i) \mathrm{d}\tau_i\\
&= \exp\left(-\sum_{i=1}^N \int_0^{\tau_i} \lambda(s)  \mathrm{d}s + \sum_{i=1}^N \mathrm{ln} (\lambda(\tau_i))\right) \prod_{i=1}^N\mathrm{d}\tau_i\\
&\simeq \exp\left( -\int_0^T \lambda(t)N(t)\mathrm{d}t + \int_0^T \mathrm{ln} (\lambda(t)) s(t)  \mathrm{d}t  \right) \prod_{i=1}^N\mathrm{d}\tau_i.
\end{aligned}
\end{equation}

\noindent The log-likelihood can therefore be written as:

\begin{equation}\label{eq:loglikelihood}
\mathrm{log}\left(\mathrm{d}\mathbb{P}\left[ s(t) \mid \lambda(t) \right] \right) = \int_0^T\left[\mathrm{ln} \left(\lambda(t)\right) s(t)  - \lambda(t)N(t) \right] \mathrm{d}t + \sum_{i=1}^N\mathrm{log}\left(\mathrm{d}\tau_i\right)
\end{equation}

\noindent Maximzing the log-likelihood therefore amounts to maximizing the integral term in the above equation. With no modelling constraints, the above expression is maximized for:

\begin{equation}
\lambda^*(t) = s(t) / N(t), 
\end{equation}

\noindent which corresponds to the de-trended skip profile (\textit{i.e.} the skip profile normalized by the number users who are still active at the time considered). We call $\lambda^*(t)$ the \emph{empirical skip intensity}. For a given model intensity $\lambda_\theta(t)$, the modelling error is therefore $\lambda_\theta(t) - \lambda^*(t)$.

\subsection{Events-responses modelling of skips}\label{sec:prior}

\begin{figure}[!t]
\includegraphics[width=0.65\textwidth]{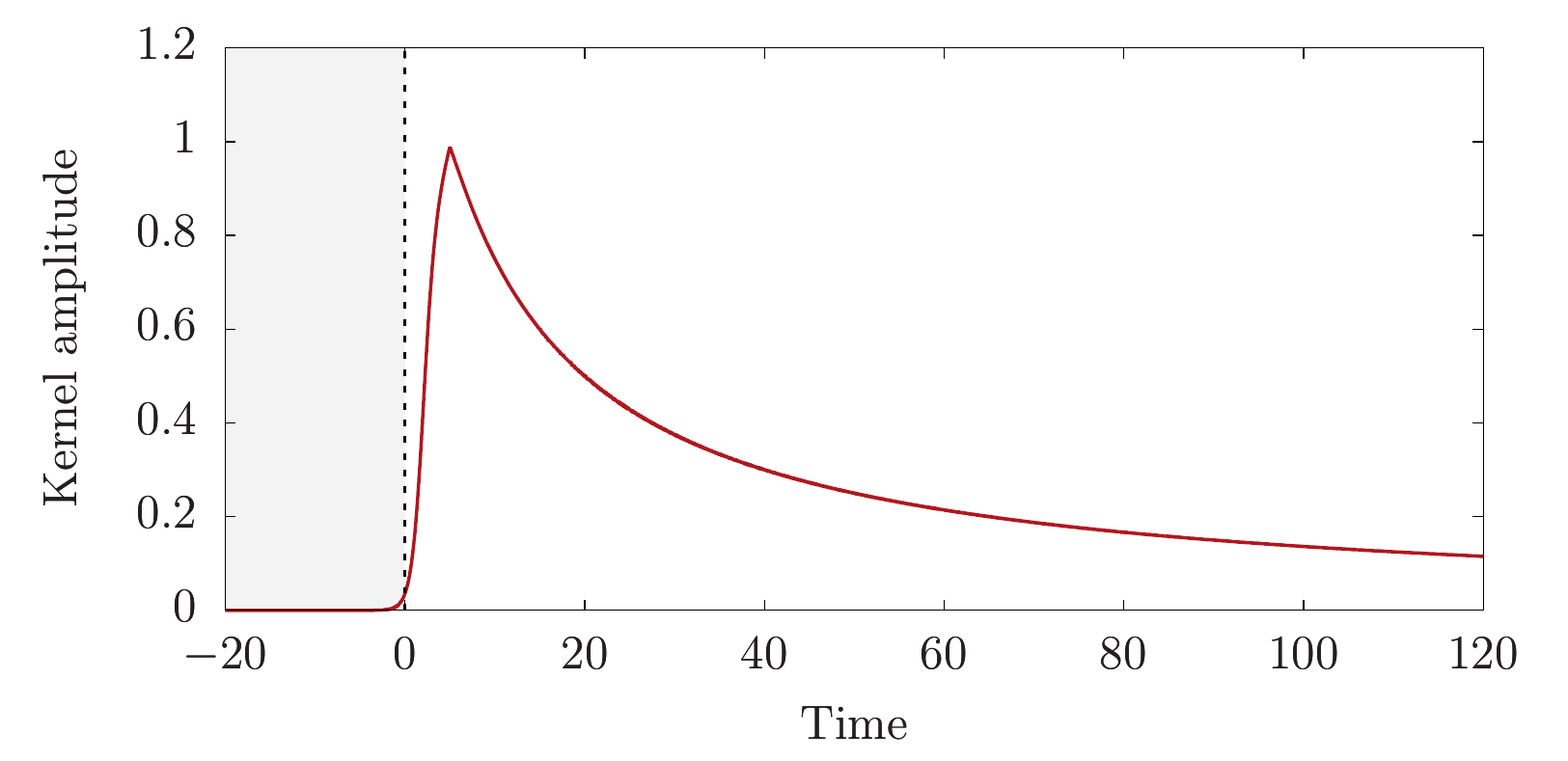}
\centering
\caption{\label{fig:kernel} The event kernel $k(t) =  \frac{1}{1 + \frac{\mid t - \tau_2 \mid}{\tau_1}}\sigma\left(\mu_\sigma(t - \tau_\sigma)\right)$, with $\tau_1 = 15$s, $\tau_2 = 5$s, $\mu_\sigma = 1.5$ and $\tau_\sigma = 2$s. The shaded region correspond to the time prior to the event, which occurs at $t=0$.}
\end{figure}

We now wish to add a prior over the model space $\mathrm{d}\mathbb{P}\left[ \lambda(t) \right]$ to reflect the structure first observed in \cite{montecchio2019skipping} and mentioned in the introduction. To this end, we model the intensity as:

\begin{equation}\label{eq:lambda}
\lambda_\theta(t) =  K_0(t) + \sum_{m=1}^M K_m(t - t_m) +  K_T(t),
\end{equation}

\noindent where:

\begin{itemize}
\item $K_0(t) := \lambda_0 / (\tau_0 + t)^{\alpha_0}$ is a power-law kernel associated with the beginning of the song, which accounts for the high initial skipping rate and the subsequent decay. 
\item $K_m(t - t_m) :=  \lambda_m k(t - t_m) $ is the kernel associated to musical event $m$, where $t_m$ is the temporal position of the event and $\lambda_m$ is the magnitude of the shock. The expression of the kernel $k$ that is used in the following sections is:

\begin{equation}
k(t) = \frac{1}{1 + \frac{\mid t - \tau_2 \mid}{\tau_1}}\sigma\left(\mu_\sigma(t - \tau_\sigma)\right),
\end{equation}

where $\sigma(\cdot)$ is the sigmoid function and the parameter values are $\tau_1 = 15$s, $\tau_2 = 5$s, $\mu_\sigma = 1.5$ and $\tau_\sigma = 2$s. As we will see in Section \ref{sec:universal}, these parameter values fit the data consistently across genres and listening contexts. Note that the initial rise happens on a time scale of $\sim 1$s, which is consistent with a recent controlled study which measured the time it takes humans to make musical aesthetic judgments \cite{belfi2018rapid}.  A graphical representation of $k(t)$ is shown on Fig. \ref{fig:kernel}.
\item $K_T(t) := \lambda_T / (\tau_T + T - t)^{\alpha_T}$ is a power-law kernel associated with the end of the song, which accounts for the increase in skipping rate as the end of the song approaches.
\end{itemize}

\noindent We assume no prior on the amplitude and exponent parameters, and consider the following prior on the temporal positions of the events:

\begin{equation}\label{eq:prior}
\mathrm{log}\left(\mathrm{d}\mathbb{P}\left[ \lambda_\theta(t) \right] \right) =  \eta \sum_{i, j=1}^M e^{-\nu\mid t_i - t_j \mid},
\end{equation}

\noindent where $\eta$ is a hyperparameter that adjusts the balance between the likelihood loss and the prior loss. This expression for the log-probability encourages sparsity for the temporal positions of events, as events that are close by are encouraged to merge into a single event. This will allow us to initialize the gradient descent with a large number of events, which will subsequently merge into a smaller number of events during the training process.\footnote{Intuitively, the prior should be high for $t_i = t_j$ but low for $t_i\simeq t_j, t_i\neq t_j$, as merged events should be encouraged while observing two transitions in a short amount of time should be penalized. However, when optimizing via gradient descent, this would lead to a potential barrier that prevents events from being merged, which is why we prefer the more regular function in Eq. (\ref{eq:prior}).} The final optimization program can therefore be written as:

\begin{equation}\label{eq:optim}
\theta^* := \underset{\theta}{\text{argmax}}~  \int_0^T\left[\mathrm{ln} \left(\lambda_\theta(t)\right) s(t)  - \lambda_\theta(t)N(t) \right] \mathrm{d}t +  \eta\sum_{i, j=1}^M e^{-\nu\mid t_i - t_j \mid},
\end{equation}

\noindent where $\theta = \{(\lambda_0, \tau_0, \alpha_0), (\lambda_m, t_m)_{m=1..M}, (\lambda_T, \tau_T, \alpha_T)\}$ and:

\begin{equation}\label{eq:lambda_explicit}
\lambda(t) =  \underbrace{\frac{\lambda_0}{(\tau_0 + t)^{\alpha_0} }}_{\text{Start kernel}}+  \sum_{m=1}^{M_0} \underbrace{\lambda_m k(t - t_m)}_{\text{Event kernels}} + \underbrace{\frac{\lambda_T}{(\tau_T + T - t)^{\alpha_T}}}_{\text{End kernel}}.
\end{equation}

\begin{figure}[!t]
\hspace*{-1cm}\includegraphics[width=0.55\textwidth]{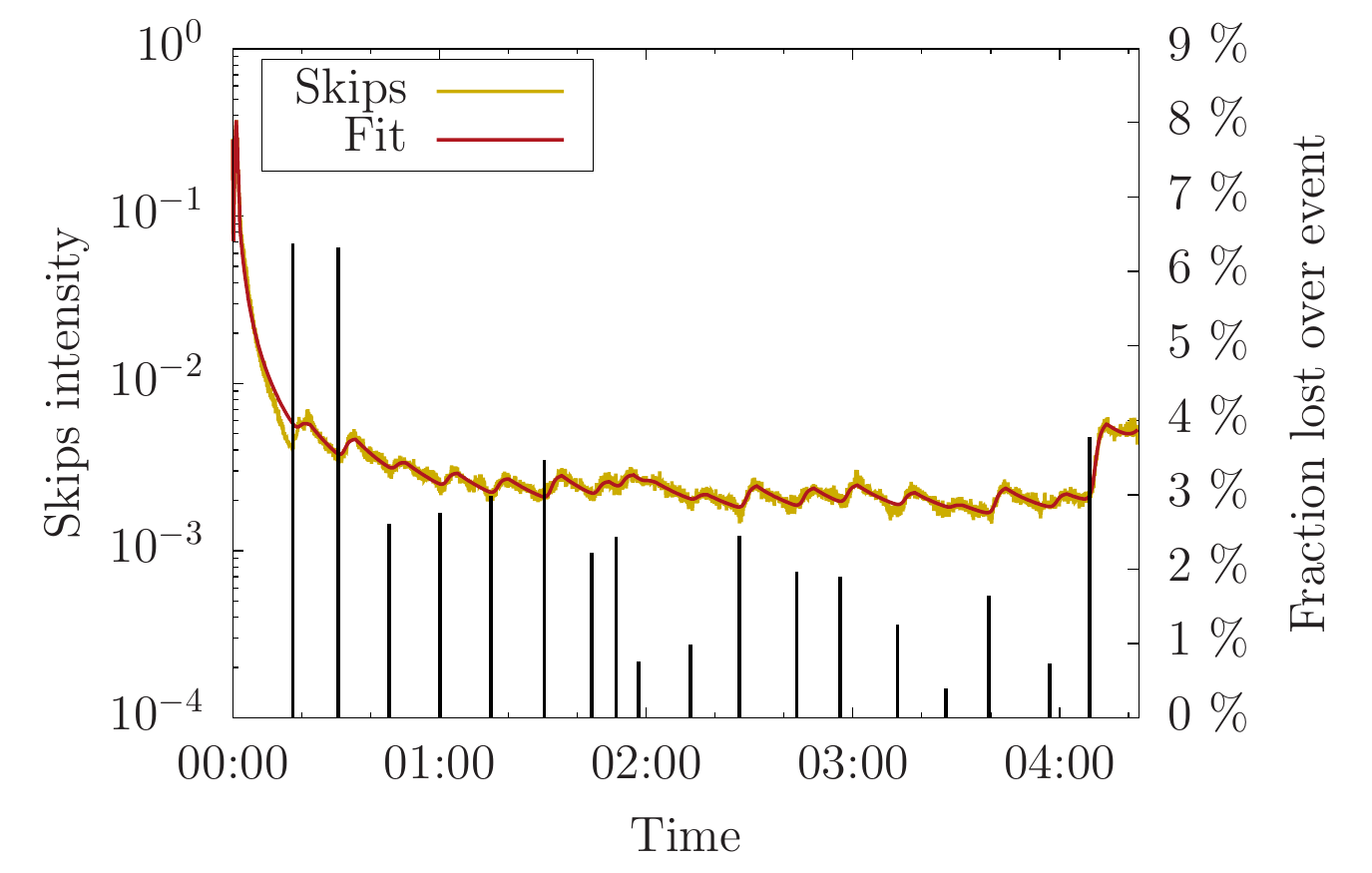}\includegraphics[width=0.55\textwidth]{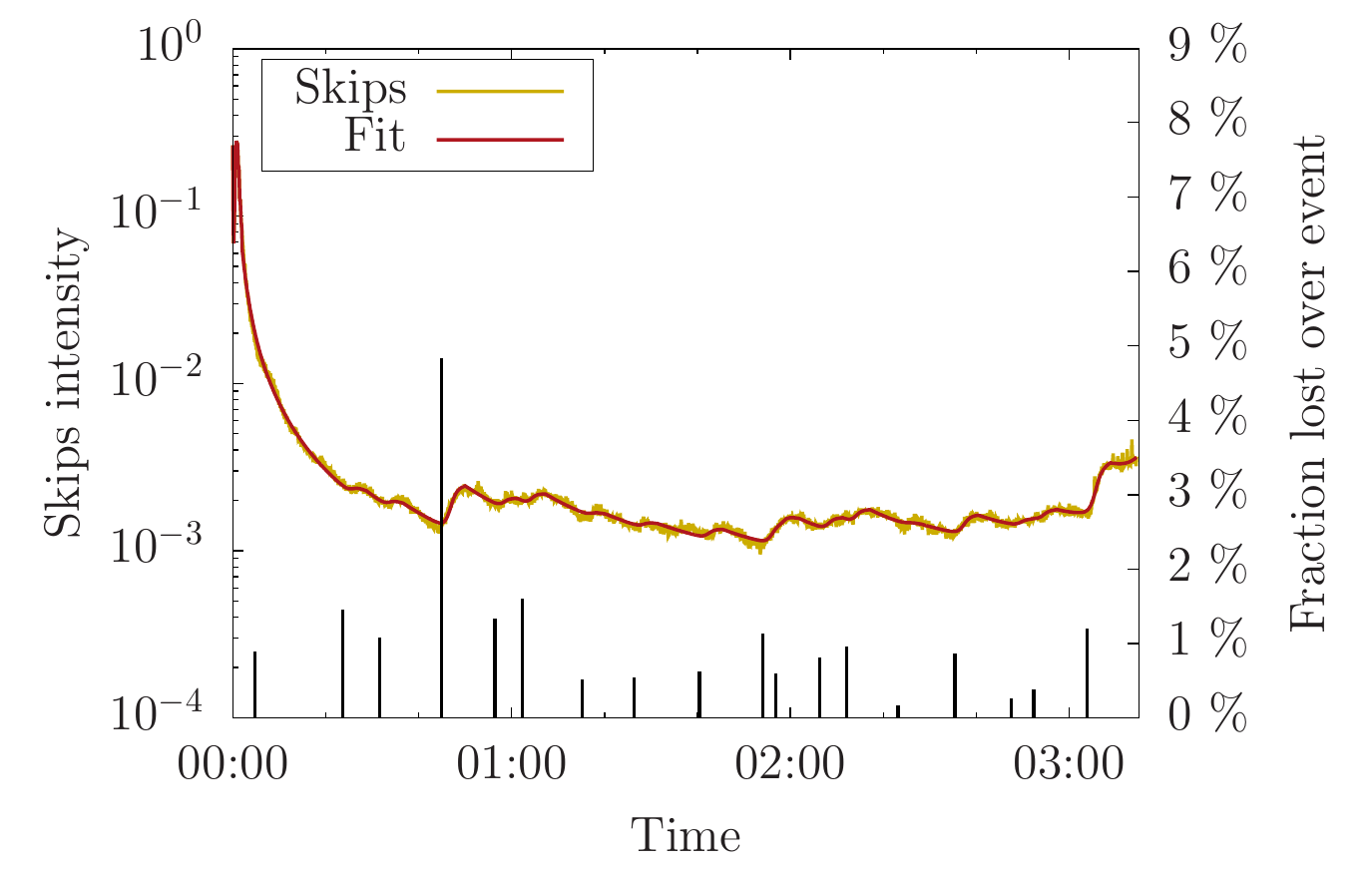}
\centering
\caption{\label{fig:intensity} Examples of empirical skip intensities $\lambda^*(t) = s(t) / N(t)$ for two different songs (in yellow) and the corresponding fitted intensities $\lambda(t)$ (in red). We use a log-scale for the $y$ axis as the skipping rate decays sharply in the first few seconds. The events $(t_m, \lambda_m)$ that are detected by the optimization process are displayed as vertical black bars whose height represent the fraction of users that have skipped because of the event (right axis).}
\end{figure}

\noindent We place the initial event times $t_m$ on a grid with 10s spacing, such that the number of events $M_0$ is larger that the actual number of events. Thanks to the shape of the prior (Eq. (\ref{eq:prior})), the $t_m$'s will gradually merge as they reach their final positions. Figure \ref{fig:intensity} shows two examples of de-trended skip profiles $s(t)/N(t)$ (i.e. the empirical Poisson intensity) and the associated model intensity $\lambda_\theta(t)$, with the events detected by the optimization process. Note that the final number of events that are detected are different for both songs, as the $t_m$'s have successfully migrated towards the true event times -- and then merged.

\section{Universality of the skips patterns}\label{sec:universal}

In this section, we wish to confirm empirically the assumptions made following Eq. (\ref{eq:lambda}) about the shapes of the various kernels.

\begin{figure}[!t]
\includegraphics[width=0.6\textwidth]{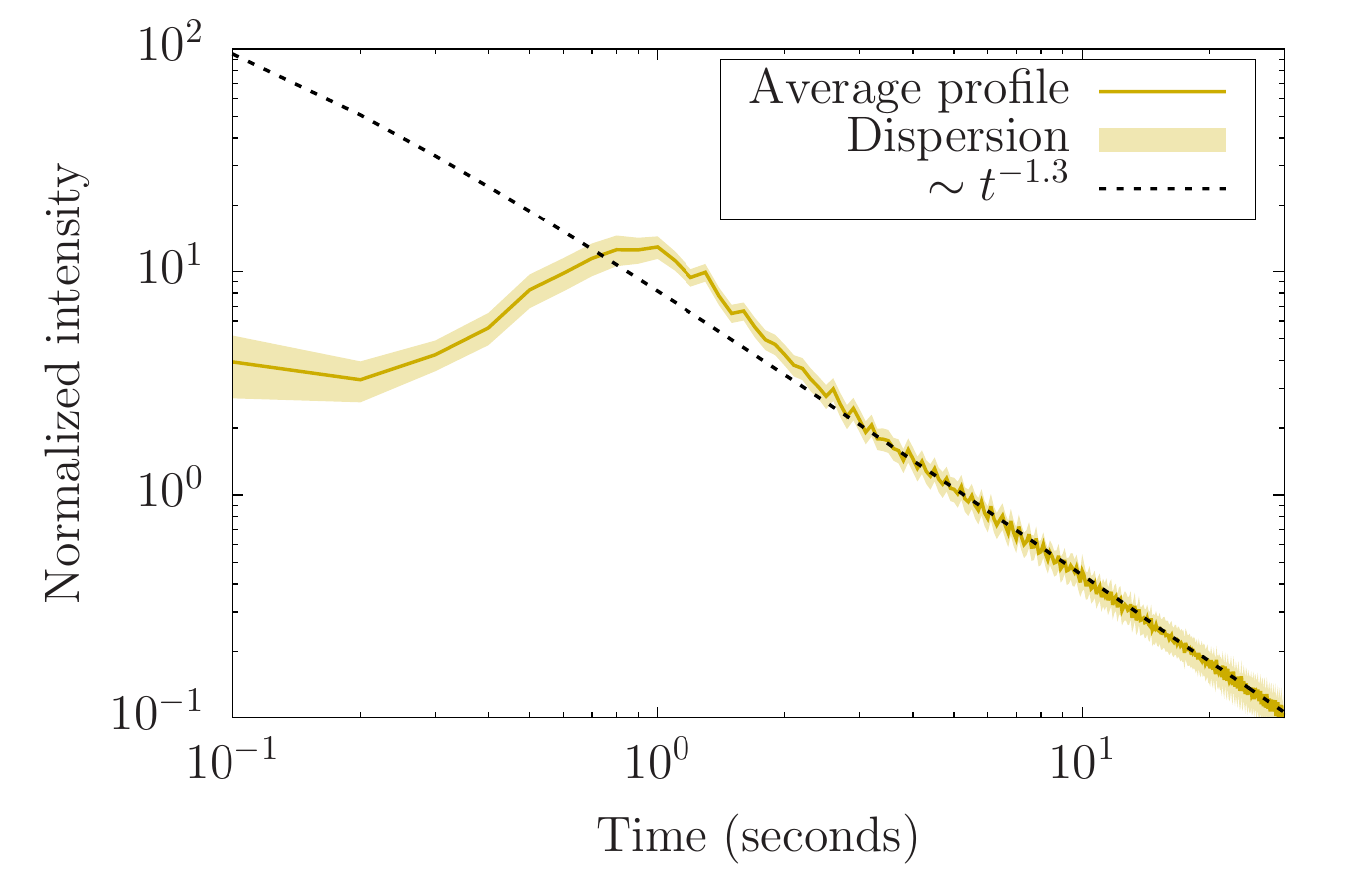}
\centering
\caption{\label{fig:decay} The normalized decay profile corresponding to the first 30s of 100 popular songs from the Spotify catalogue. The normalized mean is plotted in blue, the 1-$\sigma$ dispersion is plotted as a yellow shade. The dashed line corresponds to the initial kernel $K_0(t) := \lambda_0 / (\tau_0 + t)^{\alpha_0}$ with parameters $\alpha_0=1.3$, $\tau_0 = 0.06$s.}
\end{figure}

\paragraph{Initial decay} Fig. \ref{fig:decay} shows the normalized decay for the first 30s for 100 popular songs from the Spotify catalogue, showing a remarkable stability across tracks, with a decay exponent $\alpha_0 \simeq 1.25$ and an offset $\tau_0 \simeq 0.5$s -- thus confirming the relevance of the power-law kernel shape for the initial decay. Note that the very first seconds correspond to a rather different mechanism and deviate from the power law, so we exclude the first 2s from the fit.

\begin{figure}[!t]
\hspace*{1cm}\includegraphics[width=0.9\textwidth]{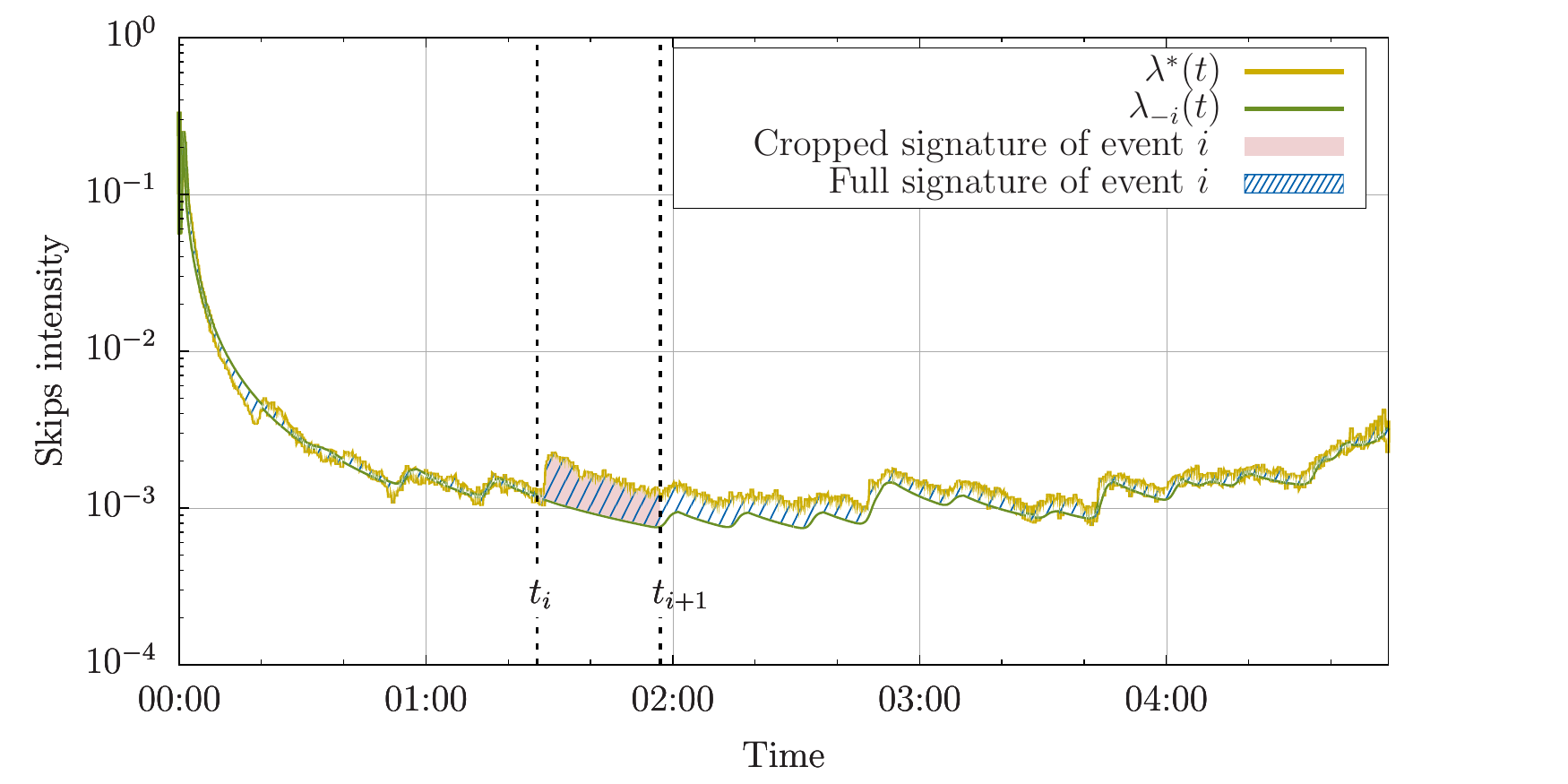}
\centering
\caption{\label{fig:diff} The empirical skip intensity $\lambda^*(t)$ (in yellow), and the model intensity from which we have removed the kernel associated with event $i$, $\lambda_{-i}(t)$ (in green). The pink area represents the signature of the event cropped at the next event $i+1$, while the hatched area represents the full signature.}
\end{figure}

\begin{figure}[!ht]
\includegraphics[width=\textwidth]{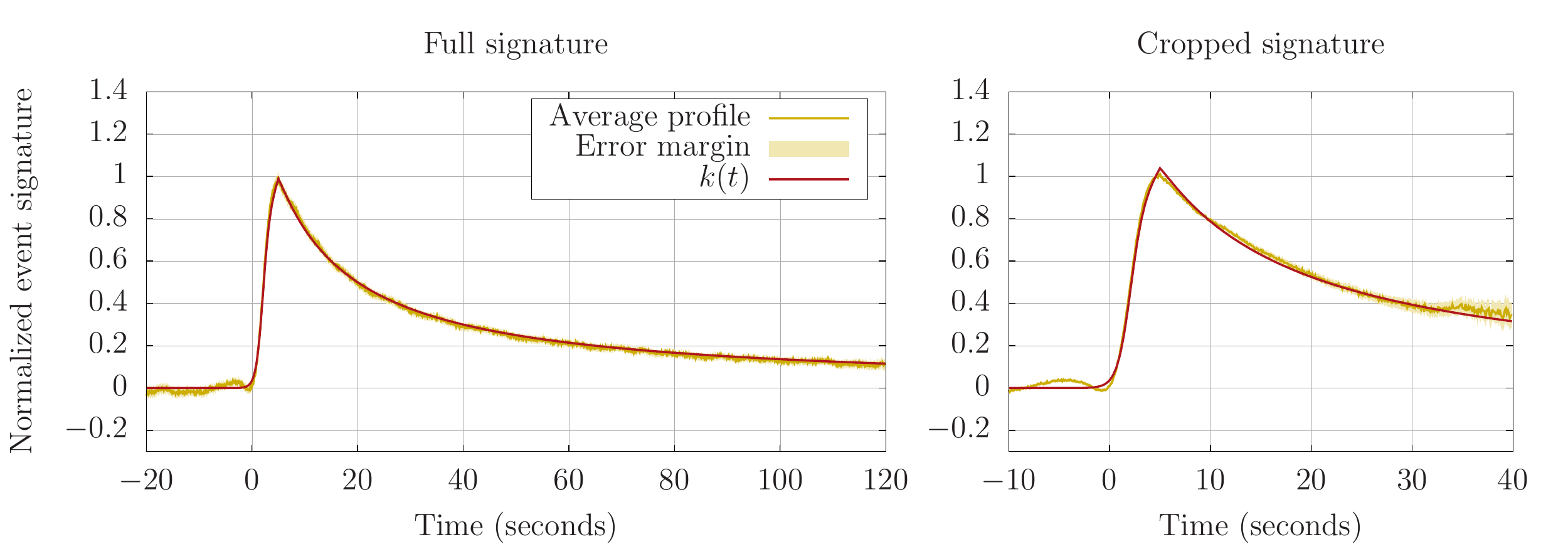}
\centering
\caption{\label{fig:kernel_diff} The average signature of events, centred around the estimated event times ($t=0$). The curves are max-normalized prior to averaging. The $\pm 3\sigma$ error margin is plotted as a yellow shade. The red line corresponds to the event kernel $k(t) =  \frac{1}{1 + \frac{\mid t - \tau_2 \mid}{\tau_1}}\sigma\left(\mu_\sigma(t - \tau_\sigma)\right)$ with parameters  $\tau_1 = 15$s, $\tau_2 = 5$s, $\mu_\sigma = 1.5$ and $\tau_\sigma = 2$s. The kernel is very close to both the full and the cropped empirical signatures and well within the statistical error bounds.}
\end{figure}

\paragraph{Spikes} We now turn our attention to the event kernel $k(t)$. For every event found by the model, we compute the difference $\lambda^*(t) - \lambda_{-i}(t)$ between the empirical skip intensity $\lambda^*(t)$ and the optimal intensity profile found by the model from which we have removed the kernel associated with this very event, defined as:

\begin{equation}\label{eq:lambda_i}
\lambda_{-i}(t) =  \frac{\lambda_0}{(\tau_0 + t)^{\alpha_0} }+  \sum_{\substack{m=1\\ m\neq i}}^{M_0} \lambda_m k(t - t_m) + \frac{\lambda_T}{(\tau_T + T - t)^{\alpha_T}}.
\end{equation}

\begin{figure}[!t]
\centering
\includegraphics[width=0.72\textwidth]{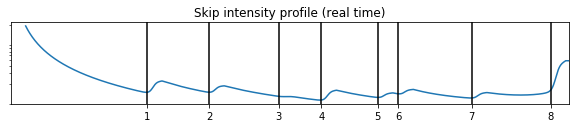}\\
\includegraphics[width=0.72\textwidth]{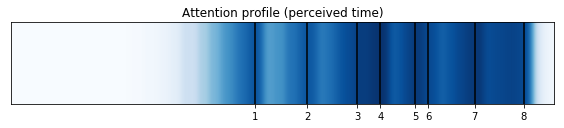}

\caption{\label{fig:attention} \textit{(top)} The intensity profile (in blue) and chronology of events (black vertical bars) in real time. \textit{(bottom)} The attention map and chronology of events (black vertical bars) in perceived time. Sections with high attention and skip rate are dilated in perceived time (lighter sections), while monotonous sections are contracted (darker sections).}
\end{figure}

\noindent This gives us the theoretical intensity corresponding to every event \emph{except} the one under consideration (cf. Figure \ref{fig:diff}). We then analyze this difference -- which we call the empirical signature of the event -- to confirm whether its shape matches that of the kernels use in our model. We analyze both the full signature (hatched area on Figure \ref{fig:diff}) and the cropped signature where we discard everything that happens after the next event (pink area on Figure \ref{fig:diff}). The advantage of the cropped signature is that it limits the influence of surrounding kernels on shape of the empirical signature: in that sense, it is closer to a ``true'' signature of the event. The results are plotted on Figure \ref{fig:kernel_diff}, showing that both the cropped and the full skip signatures match the shape of $k(t)$ with good accuracy. We show in Appendix \ref{sec:signature_cross_analysis},  Figures \ref{fig:kernel_dissection_genres} and  \ref{fig:kernel_dissection_contexts} that this remains true across genres and for various listening contexts (free or premium subscription, and whether or not the song was played based on a recommendation by the platform). Note that this does not directly prove that the chosen kernel shape is the ``true'' shape, if such shape exists: only that it provides a good basis for decomposing the skip curve. However, a similar analysis for different kernel shapes is performed in Appendix \ref{sec:comparison}, most of which are unable to reproduce the above fit quality. This confirms the relevance of the above shape for the event kernels, with a linear increase in the first few seconds after the event followed by a slow decay. This essentially shows two things:

\begin{enumerate}[(i)]
\item skips are triggered by punctual events -- or, more precisely, they almost exclusively depend on the time elapsed since the beginning of a continuous section,
\item the skip rate associated with a continuous section dies off slowly. One interpretation could be that the perception of time in music varies: the longer users have been listening to a continuous musical segment, the weaker their perception of time --  whereas abrupt transitions, on the contrary, awakens this perception. This idea is illustrated on Figure \ref{fig:attention}, which compares the chronology of events in real and perceived times -- i.e. the time in which the skip rate would be constant -- such that most of the perceived time corresponds to the very first seconds of the track where the attention (and the skip rate) are maximal, while more monotonous sections are contracted.
\end{enumerate}

\noindent This sigmoid-power-law shape is very close to a log-normal shape, which has been documented by several studies as characteristic of human response times in various contexts \cite{buzsaki2014log, schnipke1999representing, holden2009dispersion, holden2002fractal, van2003self, thissen1983timed} and which has been used extensively in subsequent response time modelling \cite{van1999using, ulrich1993information, van2006lognormal, fox2007modeling}. By confirming these results at un unprecedented scale, the present study shows that music is no exception.

\paragraph{Final increase} The pattern associated with the final increase is less clear-cut and universal than the patterns associated with the beginning of the song and with the events within the song. Since the average pattern seems to correspond to a power-law kernel with exponent $\alpha_T\simeq 0.4$, we keep a power-law kernel for modelling the final increase, as it substantially improves the overall fit of the skip curve. However, there is no clear indication of a universal pattern for the final increase.

\section{Stability of the spikes}\label{sec:spikes}

It was suggested in \cite{montecchio2019skipping} that the skip profile of a song can be seen as a fingerprint of the song, as it is highly stable across geographical regions and time periods while being highly specific to the song. In the previous sections we have shown that the skip curve can be parametrized by a very small number of parameters -- essentially, the timings $t_m$ and amplitudes $\lambda_m$ associated with a discrete number of events. In this section, we test the hypothesis that the set of parameters associated with the events $\{(t_m, \lambda_m)\}$ can be used as a compact fingerprint of a song, which would confirm the findings in \cite{montecchio2019skipping} and validate our model at the same time. We thus need to answer the following question: given two sets of $M_1$ and $M_2$ event parameters $\theta^1 = \{(t_m^1, \lambda_m^1)\}$ and $\theta^2 = \{(t_m^2, \lambda_m^2)\}$, what is the probability that they correspond to the same song? 

\subsection{Posterior probability}

We take a bayesian approach, and use the following notations:

\begin{itemize}
\item $\mathbb{P}\left[ \text{same} \mid (\theta^1, \theta^2) \right]$ is the probability that songs 1 and 2 associated with parameters $\theta^1$ and $\theta^2$ are the same song. This is the quantity we ultimately wish to compute.
\item $\mathrm{d}\mathbb{P}\left[(\theta^1, \theta^2)  \mid \text{same} \right]$ is the probability to have observes $(\theta^1, \theta^2)$ if songs 1 and 2 were the same,
\item $\mathrm{d}\mathbb{P}\left[(\theta^1, \theta^2)  \mid \text{different} \right]$ is the probability to have observes $(\theta^1, \theta^2)$ if songs 1 and 2 were different,
\item $\mathbb{P}\left[ \text{same} \right]$ is the prior probability of two songs being the same,
\item $\mathbb{P}\left[ \text{different} \right]$ is the prior probability of two songs being different.
\end{itemize}

\noindent We can then write, using Bayes rule:

\begin{equation}
\begin{aligned}
\mathbb{P}\left[ \text{same} \mid (\theta^1, \theta^2) \right] &= \left( 1 + \frac{\mathrm{d}\mathbb{P}\left[(\theta^1, \theta^2)  \mid \text{different} \right]}{\mathrm{d}\mathbb{P}\left[(\theta^1, \theta^2)  \mid \text{same} \right]} \frac{\mathbb{P}\left[ \text{different} \right]}{\mathbb{P}\left[ \text{same} \right]} \right)^{-1}\\
&= \rho\left( \frac{\mathrm{d}\mathbb{P}\left[(\theta^1, \theta^2)  \mid \text{same} \right]}{\mathrm{d}\mathbb{P}\left[(\theta^1, \theta^2)  \mid \text{different} \right]}\right),
\end{aligned}
\end{equation}

\noindent where $\rho(x) := \left(1 + n x^{-1}\right)^{-1}$ is a monotonously increasing function and we have noted $n := \frac{\mathbb{P}\left[ \text{different} \right]}{\mathbb{P}\left[ \text{same} \right]}$ the ratio between the prior probability of the two songs being different and  the prior probability of the two songs being the same (typically, $n\gg 1$). We make a naive Bayes assumption that events are independent, so that we can write:

\begin{equation}
\mathrm{d}\mathbb{P}\left[(\theta^1, \theta^2)  \mid \text{different} \right] = \Lambda^{M_1 + M_2} \prod_{m_1=1}^{M_1} h(\lambda_{m_1}) \mathrm{d}\lambda_{m_1}  \mathrm{d}t_{m_1}  \prod_{m_2=1}^{M_2}  h(\lambda_{m_2})  \mathrm{d}\lambda_{m_2}  \mathrm{d}t_{m_2},
\end{equation}

\noindent where we make the hypothesis that musical events in a song happen with a Poisson rate $\Lambda$, and we have noted $h(\lambda)$ the empirical distribution of amplitudes $\lambda$. Similarly, we make the hypothesis that the joint probability of observing $(t_{m_1}, \lambda_{m_1})$ and $(t_{m_2}, \lambda_{m_2})$ if event $m_1$ in song 1 matches event $m_2$ in song 2 can be expressed as:

\begin{equation}
\underbrace{\sqrt{h(\lambda_{m_1})h(\lambda_{m_2})}}_{\substack{\text{Amplitude of the}\\ \text{underlying event}}} \underbrace{\ell(\mid \lambda_{m_1} - \lambda_{m_2}\mid)}_{\substack{\text{Dispersion around}\\ \text{the true amplitude}}} \underbrace{f(\mid t_{m_1} - t_{m_2}\mid)}_{\substack{\text{Dispersion around}\\ \text{the true timing}}},
\end{equation}

\noindent and we write:

\begin{equation}
\begin{aligned}
\mathrm{d}\mathbb{P}\left[(\theta^1, \theta^2)  \mid \text{same} \right] =  \Lambda^{M_1 + M_2}  &\prod_{m_1 \leftrightarrow m_2} \sqrt{h(\lambda_{m_1})h(\lambda_{m_2})} \ell(\mid \lambda_{m_1} - \lambda_{m_2}\mid) \frac{ f(\mid t_{m_1} - t_{m_2}\mid)}{\Lambda} \\
\times & \prod_{m_1 \not \rightarrow m_2} h(\lambda_{m_1})  \mathbb{P}\left[ \cancel{m_2} \mid m_1 \right] \prod_{m_2 \not \rightarrow m_1} h(\lambda_{m_2})  \mathbb{P}\left[ \cancel{m_1} \mid m_2 \right]\\
\times & \prod_{m_1=1}^{M_1}\mathrm{d}\lambda_{m_1}  \mathrm{d}t_{m_1}    \prod_{m_2=1}^{M_2}\mathrm{d}\lambda_{m_2}  \mathrm{d}t_{m_2} ,
\end{aligned}
\end{equation}

\noindent with the additional notations $m_1 \leftrightarrow m_2$ for ``event $m_1$ corresponds to event $m_2$'', $m_1 \not \leftrightarrow m_2$ for ``there is no event in song 2 that corresponds to $m_1$'' (and conversely) and where $ \mathbb{P}\left[ \cancel{m_2} \mid m_1 \right]$ denotes the probability that no event is observed in song 2 if event $m_1$ is observed in song 1 (given that song 1 matches song 2). Putting the above equations together, one obtains: 

\begin{equation}\label{eq:match}
\begin{aligned}
\mathbb{P}\left[ \text{same} \mid (\theta^1, \theta^2) \right]  = \rho&\left( \prod_{m_1 \leftrightarrow m_2} \frac{\ell(\mid \lambda_{m_1} - \lambda_{m_2}\mid)}{\sqrt{h(\lambda_{m_1})h(\lambda_{m_2})}} \frac{f(\mid t_{m_1} - t_{m_2}\mid)}{\Lambda}\right.\\
&\left. \prod_{m_1 \not \rightarrow m_2} \mathbb{P}\left[ \cancel{m_2} \mid m_1 \right] \prod_{m_2 \not \rightarrow m_1}  \mathbb{P}\left[ \cancel{m_1} \mid m_2 \right] \right),
\end{aligned}
\end{equation}

\noindent where the first line inside the parenthesis gives a boost for events with matching timing and amplitudes, while the second line is a penalty for every event in one song that does not correspond to an event in the other song.

\subsection{Matching the events}

\begin{figure}[!t]
\hspace*{-1.5cm}\includegraphics[width=0.6\textwidth]{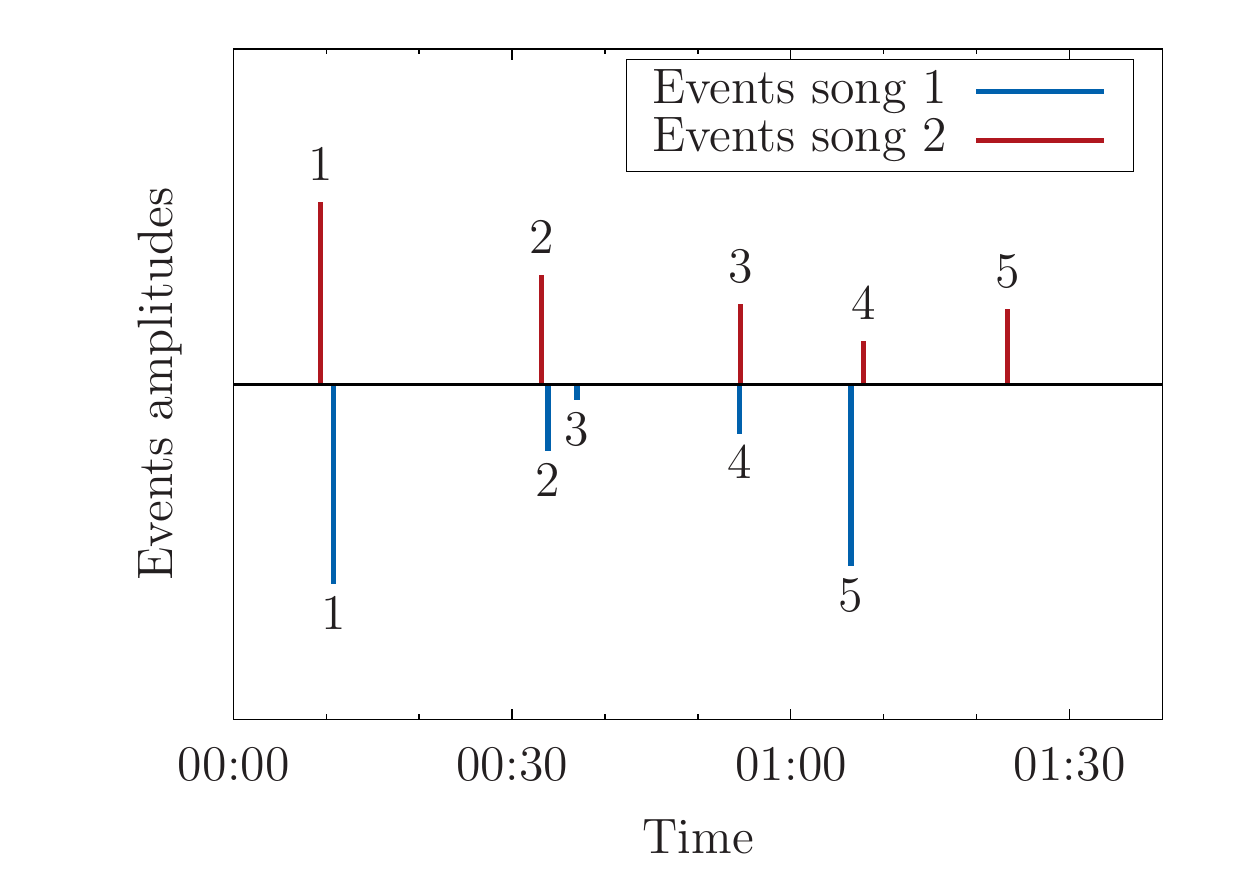} \includegraphics[width=0.47\textwidth]{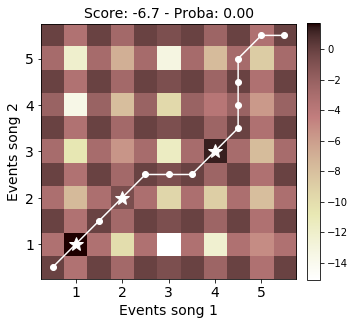}
\centering
\caption{\label{fig:grid} \textit{(left)} In blue, the timing and amplitudes of a first song (displayed on the negative $y$-axis for clarity). In red, the timing and amplitudes of a second song. \textit{(right)} The corresponding grid is shown (values are in log-scale), with the optimal path in white. Three matches are found (white stars) between pairs of events (1, 1), (2, 2) and (4, 3). The other events are matched with empty intervals (even rows/columns, represented with white circles), resulting in a low posterior probability of match between the two songs. Note that the pair (5, 4) is not matched because of the difference in amplitude.}
\end{figure}

Note that writing $m_1 \leftrightarrow m_2$ and $m_1 \not \rightarrow m_2$ assumes that events from song 1 and 2 have previously been matched. In this section, we propose to find the optimal alignment between $ \theta^1 = \{(t_m^1, \lambda_m^1)\}_{m=1\dots M_1}$ and $\theta^2 = \{(t_m^2, \lambda_m^2)\}_{m=1\dots M_2}$ that maximizes $\mathbb{P}\left[ \text{same} \mid (\theta^1, \theta^2) \right]$. We approach this maximization problem by noticing that the different configurations for the product in Eq. (\ref{eq:match}) can be viewed as the paths from the bottom left corner to the top right corner of a grid of shape $(2 M_1 + 1, 2 M_2 + 1)$, where the even rows (resp. columns) correspond to the intervals between the events in song 1 (resp. song 2) and the odd rows (resp.columns) correspond to the events in song 1 (resp. song 2). A given path therefore matches the $m_1$-th event in song 1 with the $m_2$-th event in song 2 ($m_1 \leftrightarrow m_2$) if it goes through the position $(2 m_1 + 1, 2 m_2 + 1)$. If on the contrary the $m_1$-th row is matched to an even index (corresponding to an interval), then $m_1 \not \rightarrow m_2$. The values associated with each position on the grid correspond to the factors in Eq. (\ref{eq:match}), namely:

\begin{itemize}
\item the value $\frac{\ell(\mid \lambda_{m_1} - \lambda_{m_2}\mid)}{\sqrt{h(\lambda_{m_1})h(\lambda_{m_2})}} \frac{f(\mid t_{m_1} - t_{m_2}\mid)}{\Lambda}$ is associated to the intersection of row $2 m_1 + 1$ and column $2 m_2 + 1$,
\item  the value $\mathbb{P}\left[ \cancel{m_2} \mid m_1 \right]$ (resp. $\mathbb{P}\left[ \cancel{m_1} \mid m_2 \right]$) is associated  to the intersections of row $2 m_1 + 1$ and columns $2 m_2$ for all $m_2$'s (resp. column $2 m_2 + 1$ and rows $2 m_1$ for all $m_1$'s),
\item the value 0 is associated  to the intersections of all even rows and columns.
\end{itemize}

\noindent The optimization problem can be solved by dynamic programming, by considering all the paths $\{i_k, j_k\}_k$ from $(0, 0)$ to $(2 M_1, 2 M_2)$ that satisfy $i_{k + 1} = i_{k} + 1$ if  $i_{k}$ is odd (a punctual event can only have one counterpart) and $i_{k + 1} \in \{i_{k}, i_{k} + 1\}$ if  $i_{k}$ is even (an interval can have several counterparts) -- and similarly for $j_k$. An example of such grid for two non-matching songs with the associated optimal path is represented on Figure \ref{fig:grid}. The optimal product value to use in Eq. (\ref{eq:match}) is then the product of the values along the optimal path.

\subsection{Results}

\begin{figure}[!t]
\includegraphics[width=\textwidth]{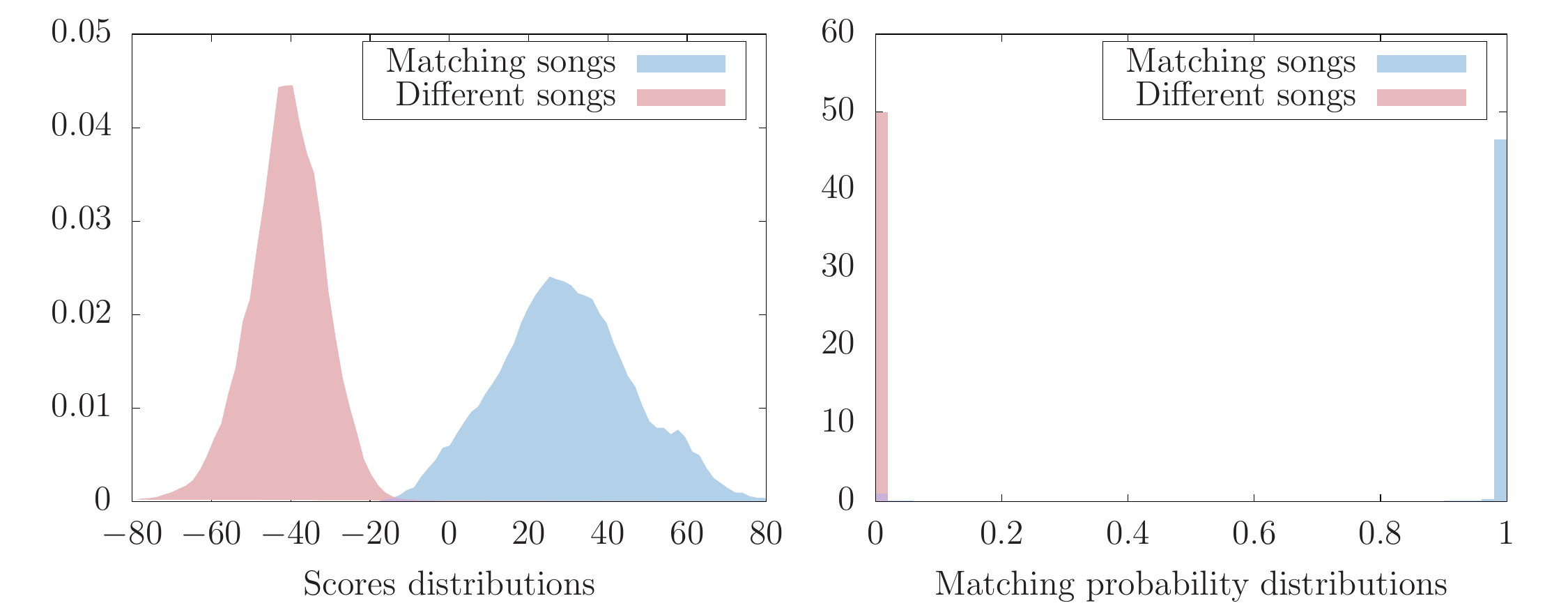}
\centering
\caption{\label{fig:matching_hist} \textit{(left)} In blue, the histogram of scores (i.e. the logarithm of the argument of $\rho(\cdot)$ in Eq. (\ref{eq:match})) corresponding to the same track on different days. In red, the histogram of scores for different tracks.  \textit{(right)} The corresponding probabilities according to Eq. (\ref{eq:match}) for an arbitrary prior $n=1$. Most non-matching tracks have a matching probability close to 0 while most matching tracks have a matching probability close to 1.}
\end{figure}

We compute the matching probabilities for the daily skip curves of a set of matching songs and a set of different songs. The log-products (also called the \textit{scores}) and the corresponding probabilities with an arbitrary prior $n = 1$ are represented on Figure \ref{fig:matching_hist}. The distributions for matching and non-matching tracks are clearly distinct, showing that the set of timing/amplitude parameters $\theta = \{(t_m, \lambda_m)\}$ are both highly specific to a song and stable across time, and thereby confirming the relevance and stability of our model. Note that one could use this result to fine-tune the model hyperparameters (in particular the weight of the prior loss $\eta$) by maximizing some distance between the two distributions.

\section*{Conclusion}

We have developed a simple yet powerful model for skipping behaviour, in which skips follow a Poisson process with time-varying intensity modelled as a sum of temporal responses (kernels), which are triggered by discrete events that happen inside the song, namely: (i) the beginning of the song, which triggers a power-law decaying kernel and usually accounts for the majority of skips, (ii) musical events (typically transitions) within the song, which each trigger a quick increase in skipping rate followed by a slow decay, and (iii) the end of the song, which is anticipated with a rise in skips. This events-responses decomposition of skip profiles provides us with a natural framework to quantitatively assess the impact of musical events on listening behaviour, and confirms the idea developed in \cite{montecchio2019skipping} that skips are for the most part reactions to salient musical events. This suggests that the perception of time when listening to music is highly dependent on the variety of the music, as though users were progressively anaesthetized by long monotonous sections and abruptly awoken by unexpected events. Moreover, the temporal profile of these reactions appears to be consistent across songs, suggesting some universal way for humans to react to musical surprises. The stability across time and geographical areas of the magnitude of the reactions to specific events further suggests that it should be possible to understand \emph{what} in the music motivates people to skip -- and to which extent. We leave this question for future work.

\section*{Acknowledgements} The author would like to thank Nicola Montecchio for helping with gathering the data, Pierre Roy for insightful discussions and Fran\c{c}ois Pachet for his overall support.

\bibliography{skips}{}
\bibliographystyle{plain}

\clearpage

\appendix

\section{Qualitative analysis of kernel shapes}\label{sec:comparison}

We proceed to a qualitative comparison of several event kernel shapes $k(t - t_m)$ (cf. Eq. (\ref{eq:lambda_explicit})). As in Section \ref{sec:universal}, we are interested in the empirical event signatures $\lambda^*(t) - \lambda_{-i}(t)$ (as defined in Eq. \ref{eq:lambda_i}) that are obtained for each kernel shape. We will look both at the full signature and the cropped signature (which is less biased by future events, so more reliable in theory). As already mentioned, that the empirical signatures correspond to the kernel shape only proves that the kernel provides a good decomposition basis for the skip curve (i.e. that the model is consistent), not that kernel is the ``true'' one (if it exists). However, a mismatch shows that the kernel shape is not even a good decomposition basis, and can be ruled out as the ``true'' shape. The full and cropped empirical signatures for 8 different kernel shapes are shown on Figure \ref{fig:shapes_comparison} below. A basic qualitative analysis of the results follows.

\begin{figure}[!b]
\centering
\begin{subfigure}[b]{\textwidth}
\centering
\hspace*{-1cm}\includegraphics[width=1.12\textwidth]{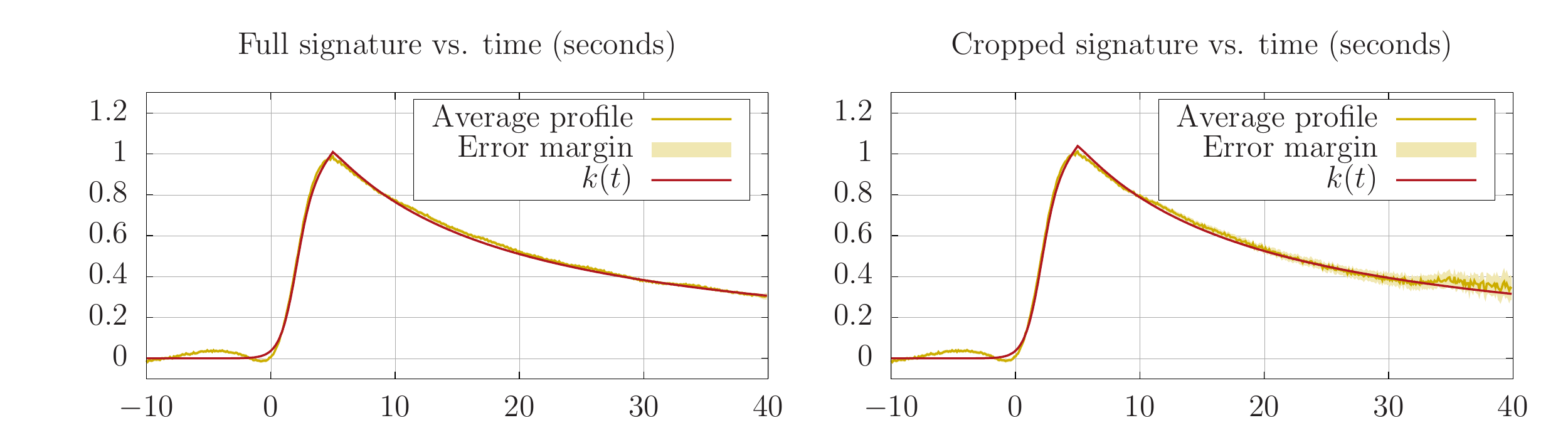}  
\caption{\label{fig:inverse_power_kernel} Inverse kernel.}
\end{subfigure}
\begin{subfigure}[b]{\textwidth}
\centering
\hspace*{-1cm}\includegraphics[width=1.12\textwidth]{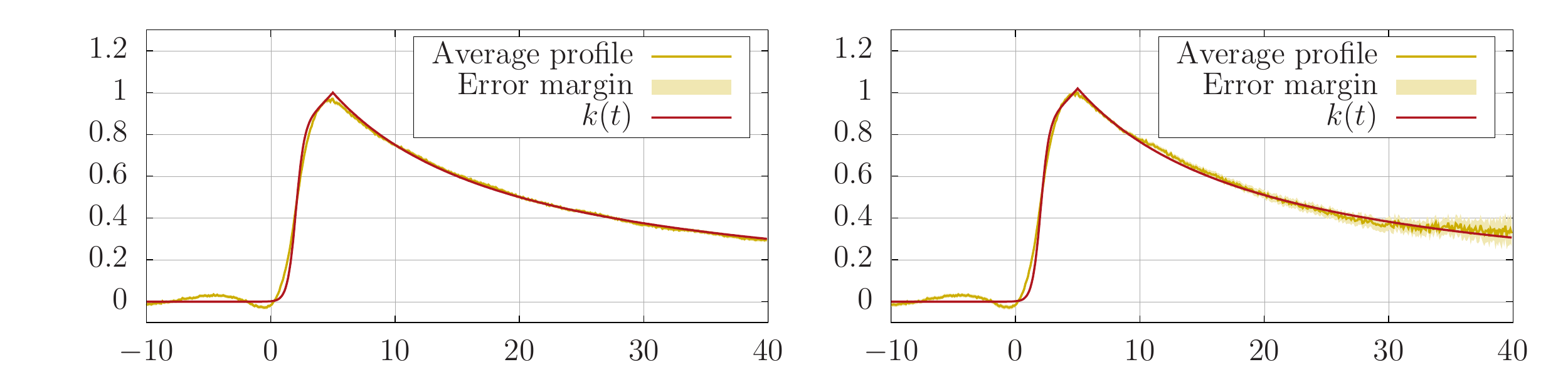}  
\caption{\label{fig:inverse_power_kernel_sharp} Inverse kernel with sharp onset.}
\end{subfigure}
\begin{subfigure}[b]{\textwidth}
\centering
\hspace*{-1cm}\includegraphics[width=1.12\textwidth]{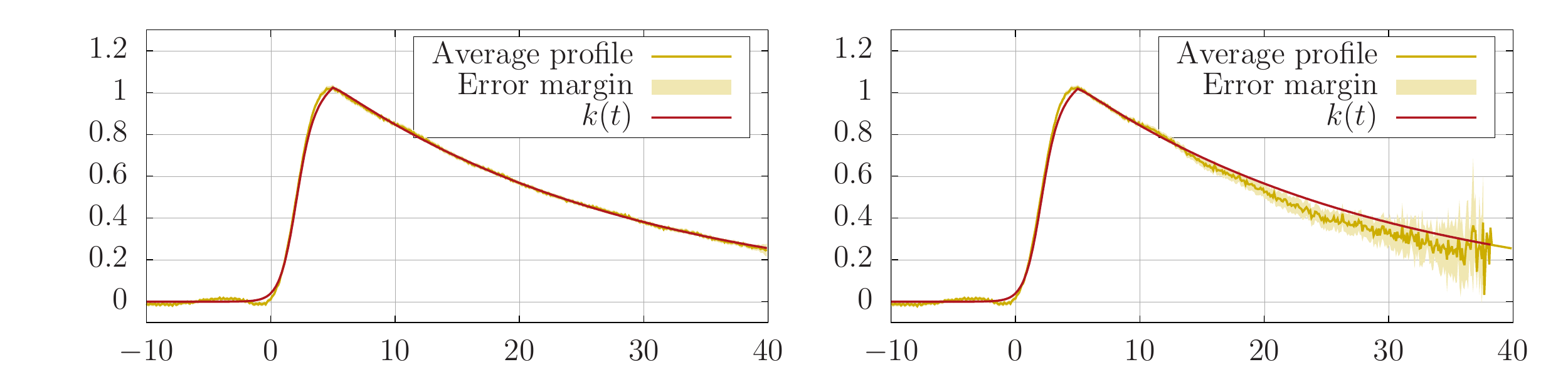}  
\caption{\label{fig:exp_kernel_004} Exponential kernel with 25s decay.}
\end{subfigure}
\end{figure}

\begin{figure}[!b]\ContinuedFloat
\begin{subfigure}[b]{\textwidth}
\centering
\hspace*{-1cm}\includegraphics[width=1.12\textwidth]{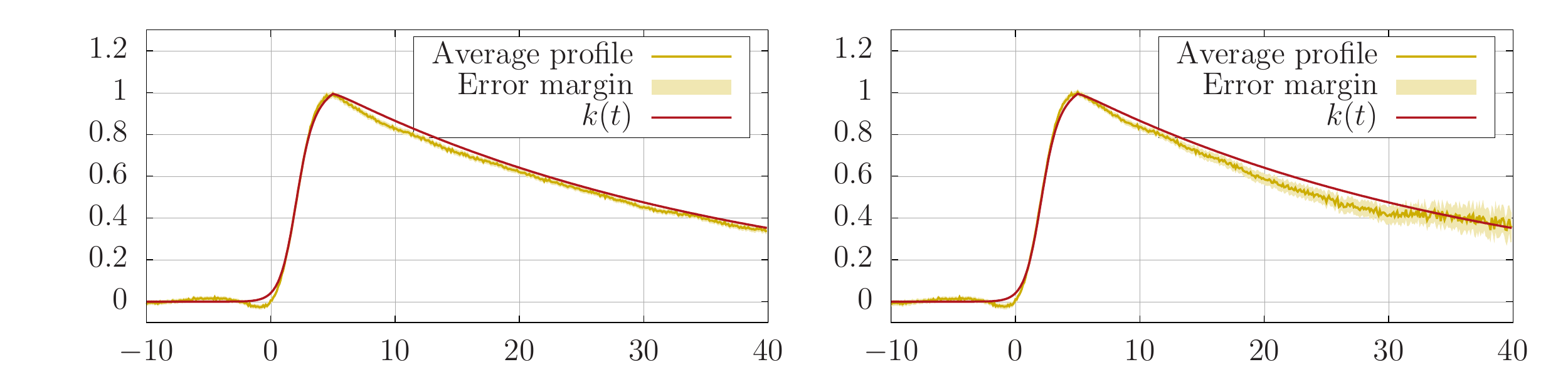}  
\caption{\label{fig:exp_kernel_003} Exponential kernel with 33s decay.}
\end{subfigure}
\begin{subfigure}[b]{\textwidth}
\centering
\hspace*{-1cm}\includegraphics[width=1.12\textwidth]{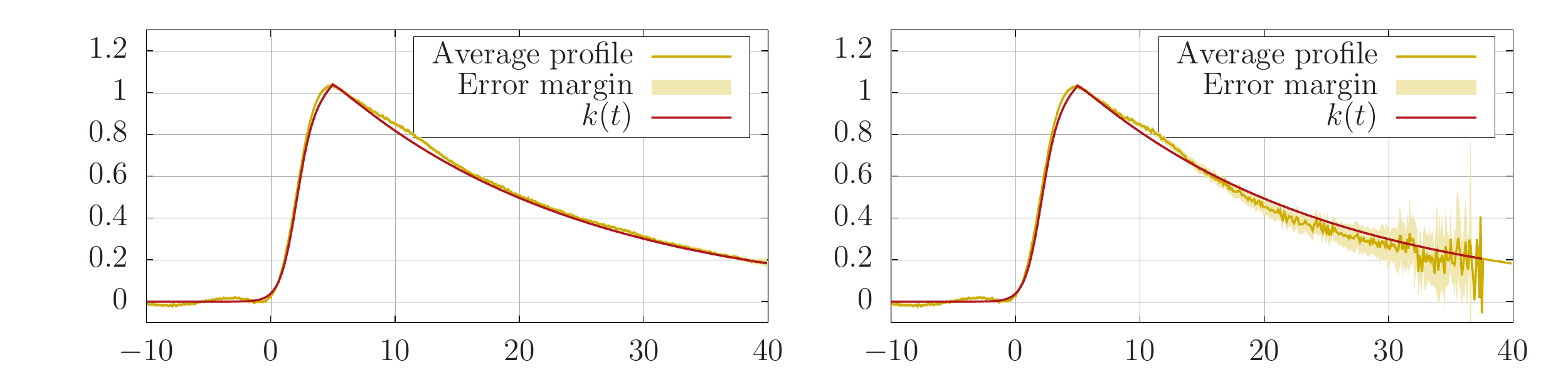}  
\caption{\label{fig:exp_kernel_005} Exponential kernel with 20s decay.}
\end{subfigure}
\begin{subfigure}[b]{\textwidth}
\centering
\hspace*{-1cm}\includegraphics[width=1.12\textwidth]{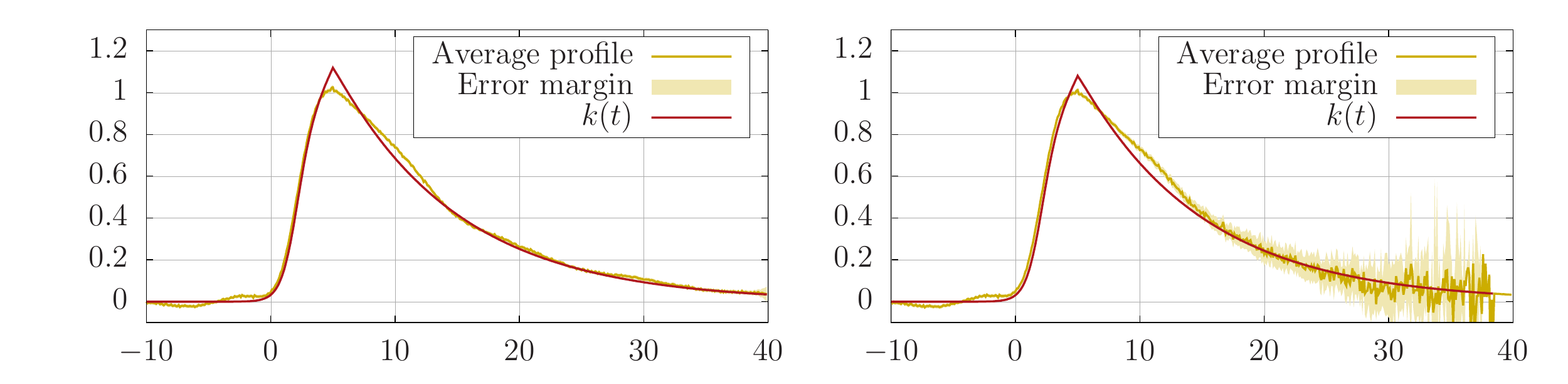}  
\caption{\label{fig:exp_kernel_010} Exponential kernel with 10s decay.}
\end{subfigure}
\begin{subfigure}[b]{\textwidth}
\centering
\hspace*{-1cm}\includegraphics[width=1.12\textwidth]{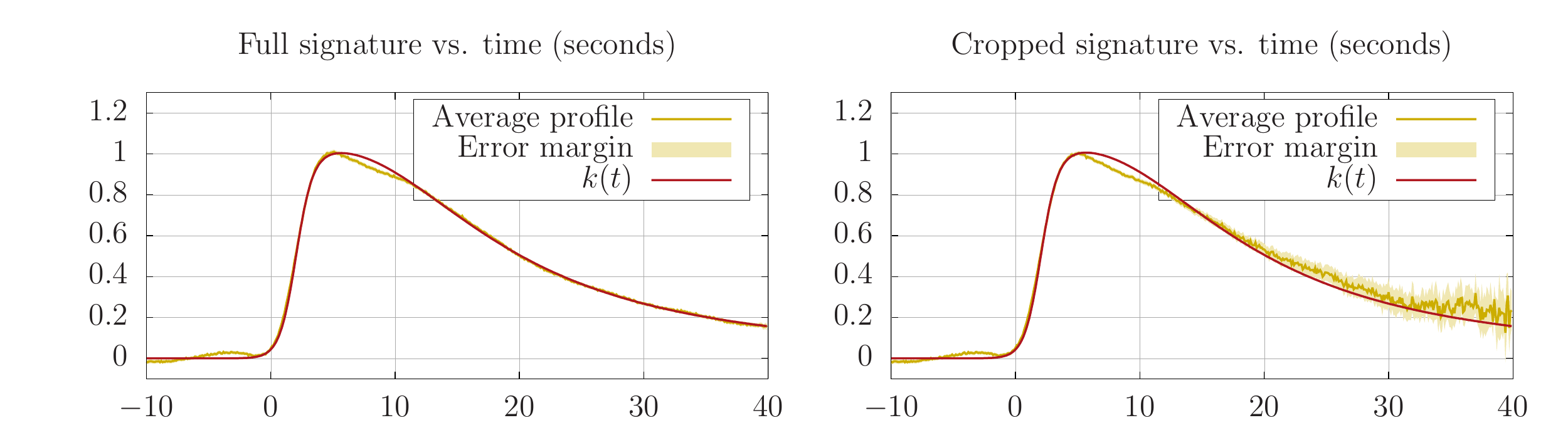}  
\caption{\label{fig:inverse_power_2b} Inverse square kernel \#1.}
\end{subfigure}
\end{figure}

\clearpage
\clearpage

\begin{figure}[!t]\ContinuedFloat
\begin{subfigure}[b]{\textwidth}
\centering
\hspace*{-1cm}\includegraphics[width=1.11\textwidth]{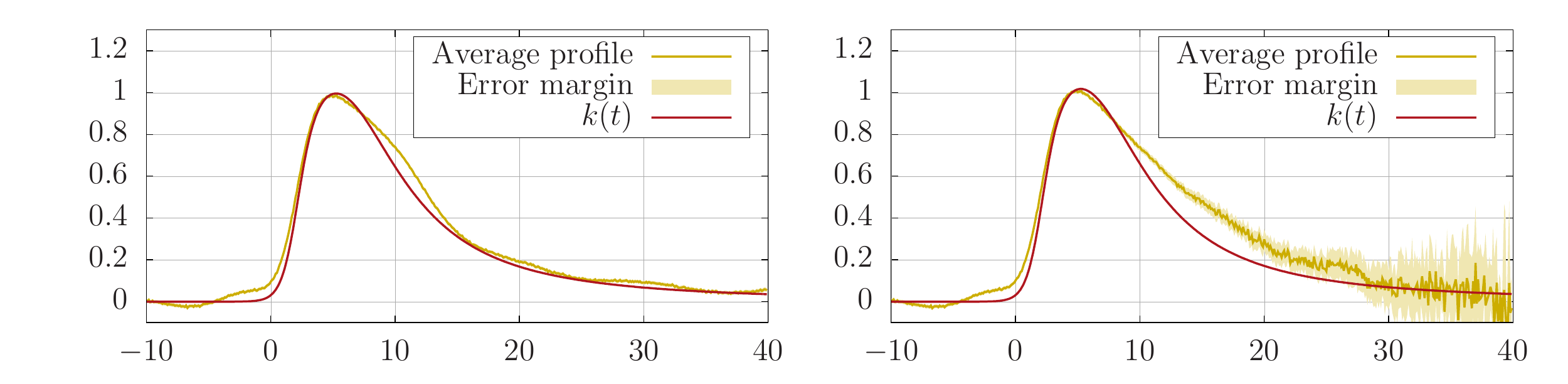}  
\caption{\label{fig:inverse_power_2} Inverse square kernel \#2.}
\end{subfigure}
\caption{\label{fig:shapes_comparison}A comparison of kernel shapes with the corresponding event signatures.}
\end{figure}

\begin{enumerate}
\item \textbf{Figure \ref{fig:inverse_power_kernel}:} Inverse kernel.
	\begin{enumerate}[(i)]
	\item Kernel shape: $k(t) = k(t) = \frac{1}{1 + \frac{\mid t - \tau_2 \mid}{\tau_1}}\sigma\left(\mu_\sigma(t - \tau_\sigma)\right)$.
	\item Parameters: $\tau_1 = 15$s, $\tau_2 = 5$s, $\mu_\sigma = 1.5$, $\tau_\sigma = 2$s.
	\item Comments: this is the kernel used in the main text. The full and cropped empirical signatures match the kernel, which thus provides a good basis for decomposition.
	\end{enumerate}
\item \textbf{Figure \ref{fig:inverse_power_kernel_sharp}:} Inverse kernel with sharp onset.
	\begin{enumerate}[(i)]
	\item Kernel shape: $k(t) = k(t) = \frac{1}{1 + \frac{\mid t - \tau_2 \mid}{\tau_1}}\sigma\left(\mu_\sigma(t - \tau_\sigma)\right)$.
	\item Parameters: $\tau_1 = 15$s, $\tau_2 = 5$s, $\mu_\sigma = 3$, $\tau_\sigma = 2$s.
	\item Comments: this kernel is similar to the inverse power kernel above, albeit with a sharper initial onset. The empirical signatures show a slower initial onset than the kernel, which would be better fitted by the previous inverse kernel with $\mu_\sigma = 1.5$. This confirms that the previous kernel better accounts for the initial onset.
	\end{enumerate}
\item \textbf{Figure \ref{fig:exp_kernel_004}:} Exponential kernel with 25 decay
	\begin{enumerate}[(i)]
	\item Kernel shape: $k(t) = e^{- \frac{\mid t - \tau_2 \mid}{\tau_1}}\sigma\left(\mu_\sigma(t - \tau_\sigma)\right)$.
	\item Parameters:  $\tau_1 = 25$s, $\tau_2 = 5$s, $\mu_\sigma = 1.5$, $\tau_\sigma = 2$s.
	\item Comments: This kernel is similar to the inverse power kernel but with a quicker long-term decay. The quality of fit is quite similar to that of the inverse power kernel. We can see by looking at the cropped empirical signature that the measurement uncertainty gets large after 30-40s, which makes it difficult to compare the fit quality with that of the inverse kernel at these time scales.
	\end{enumerate}
\item \textbf{Figure \ref{fig:exp_kernel_003}:} Exponential kernel with 33s decay.
	\begin{enumerate}[(i)]
	\item Kernel shape: $k(t) = e^{- \frac{\mid t - \tau_2 \mid}{\tau_1}}\sigma\left(\mu_\sigma(t - \tau_\sigma)\right)$.
	\item Parameters:  $\tau_1 = 33$s, $\tau_2 = 5$s, $\mu_\sigma = 1.5$, $\tau_\sigma = 2$s.
	\item Comments: A slower decay lead to a poorer fit than with the previous exponential kernel. In particular, the empirical signatures decay faster than the kernel after 5s. This is consistent with the fact that this kernel decays too slowly in the first 30s.
	\end{enumerate}
\item \textbf{Figure \ref{fig:exp_kernel_005}:} Exponential kernel with 20s decay.
	\begin{enumerate}[(i)]
	\item Kernel shape: $k(t) = e^{- \frac{\mid t - \tau_2 \mid}{\tau_1}}\sigma\left(\mu_\sigma(t - \tau_\sigma)\right)$.
	\item Parameters:  $\tau_1 = 20$s, $\tau_2 = 5$s, $\mu_\sigma = 1.5$, $\tau_\sigma = 2$s.
	\item Comments: We observe the inverse phenomenon from the slower exponential kernel, with an empirical signature that decays more slowly than the kernel around 10-15s. This is consistent with the fact that this kernel decays too quickly, in particular in the first seconds.
	\end{enumerate}
\item \textbf{Figure \ref{fig:exp_kernel_010}:}  Exponential kernel with 10s decay.
	\begin{enumerate}[(i)]
	\item Kernel shape: $k(t) = e^{- \frac{\mid t - \tau_2 \mid}{\tau_1}}\sigma\left(\mu_\sigma(t - \tau_\sigma)\right)$.
	\item Parameters:  $\tau_1 = 10$s, $\tau_2 = 5$s, $\mu_\sigma = 1.5$, $\tau_\sigma = 2$s.
	\item Comments: We observe the same phenomenon than for the previous kernel but much more accentuated. Now the empirical signature is largely out of the statistical error band.
	\end{enumerate}
\item \textbf{Figure \ref{fig:inverse_power_2b}:}  Inverse square kernel \#1.
	\begin{enumerate}[(i)]
	\item Kernel shape: $k(t) = k(t) = \frac{1}{1 + \left(\frac{ t - \tau_2 }{\tau_1}\right)^2}\sigma\left(\mu_\sigma(t - \tau_\sigma)\right)$.
	\item Parameters: $\tau_1 = 15$s, $\tau_2 = 5$s, $\mu_\sigma = 1.5$, $\tau_\sigma = 2$s.
	\item Comments: The shape of the kernel poorly matches the empirical signature around 5-10s. The crossover between the onset and the decay isn't properly captured.
	\end{enumerate}
\item \textbf{Figure \ref{fig:inverse_power_2}:}  Inverse square kernel \#2.
	\begin{enumerate}[(i)]
	\item Kernel shape: $k(t) = k(t) = \frac{1}{1 + \left(\frac{ t - \tau_2 }{\tau_1}\right)^2}\sigma\left(\mu_\sigma(t - \tau_\sigma)\right)$.
	\item Parameters: $\tau_1 = 7$s, $\tau_2 = 5$s, $\mu_\sigma = 1.5$, $\tau_\sigma = 2$s.
	\item Comments: Same as above, with a faster decay. Poor fit overall.
	\end{enumerate}
\end{enumerate}

\noindent Because of the amount of computations required to test each kernel, we have been limited in our comparative study. However, we can draw a few conclusions from these experiments:

\begin{itemize}
\item Not all kernels are consistent for modelling the skip curves. In fact, most of the kernels produce statistically significant biases in the empirical signatures.
\item It is possible to link the kernel shapes with the biases observed. In particular, one can tell how to modify the kernel onset to better fit the data, and whether the following decay is too fast or too slow.
\item After 30-40s, the statistical error bars widen, and it becomes impossible to discriminate between reasonably-shaped kernels.
\end{itemize}

We can conclude that the ``true'' kernel, if it exists, resembles the inverse power kernel (1) and the exponential kernel (3) for the first 30-40s. Whether the decay is faster or slower after that is impossible to settle. Importantly, these results remain valid across various partitions of the data (genres, listening contexts, etc.). This shows that there is some universality in the temporal response to musical events.

\section{Analysis of the empirical signature across genres and contexts}\label{sec:signature_cross_analysis}

\begin{figure}[!h]
\centering
\hspace*{-1cm}\includegraphics[width=1.1\textwidth]{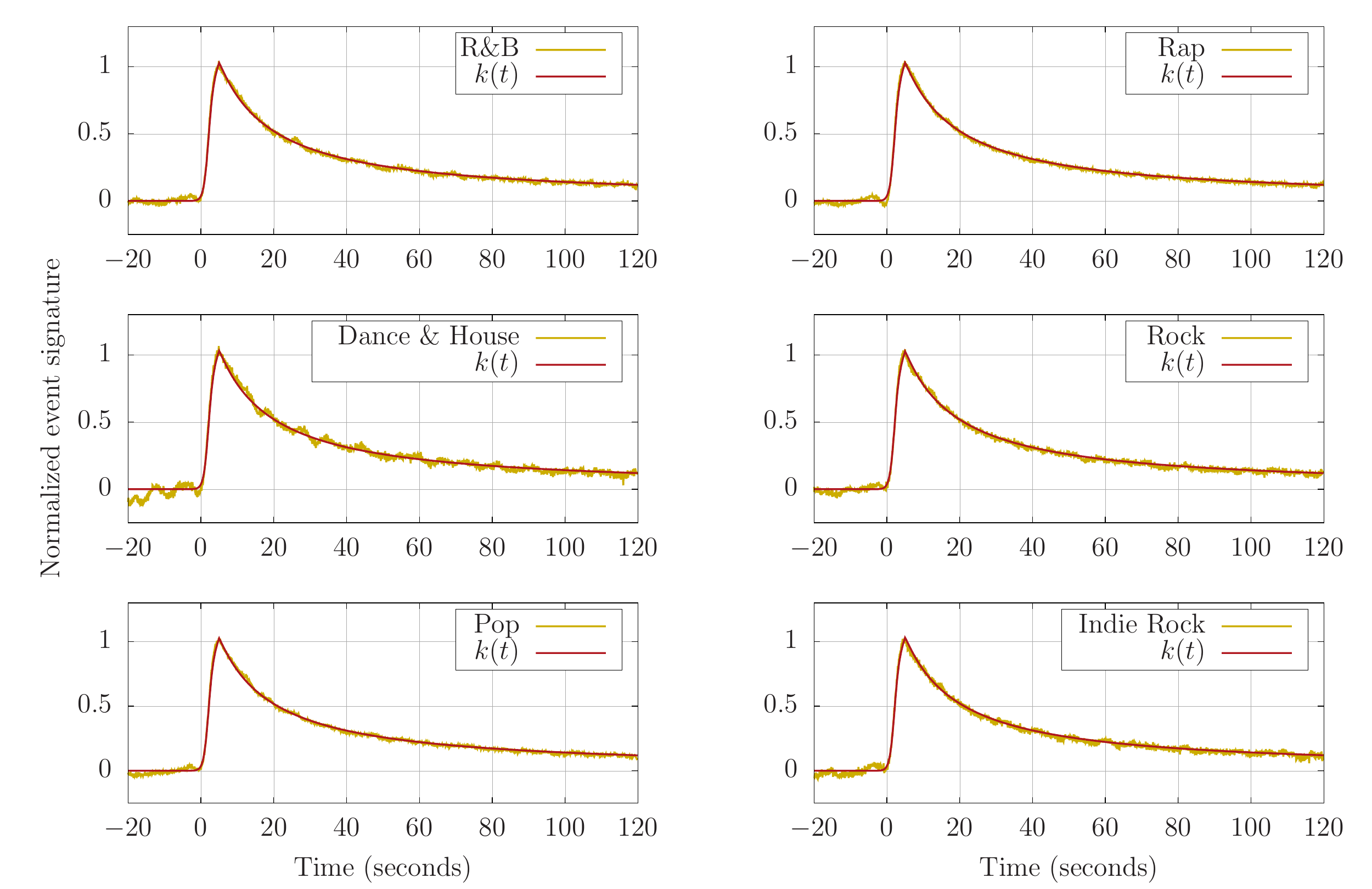}
\caption{\label{fig:kernel_dissection_genres} Average event signature for a number of musical genres.}
\end{figure}

\begin{figure}[!h]
\centering
\includegraphics[width=\textwidth]{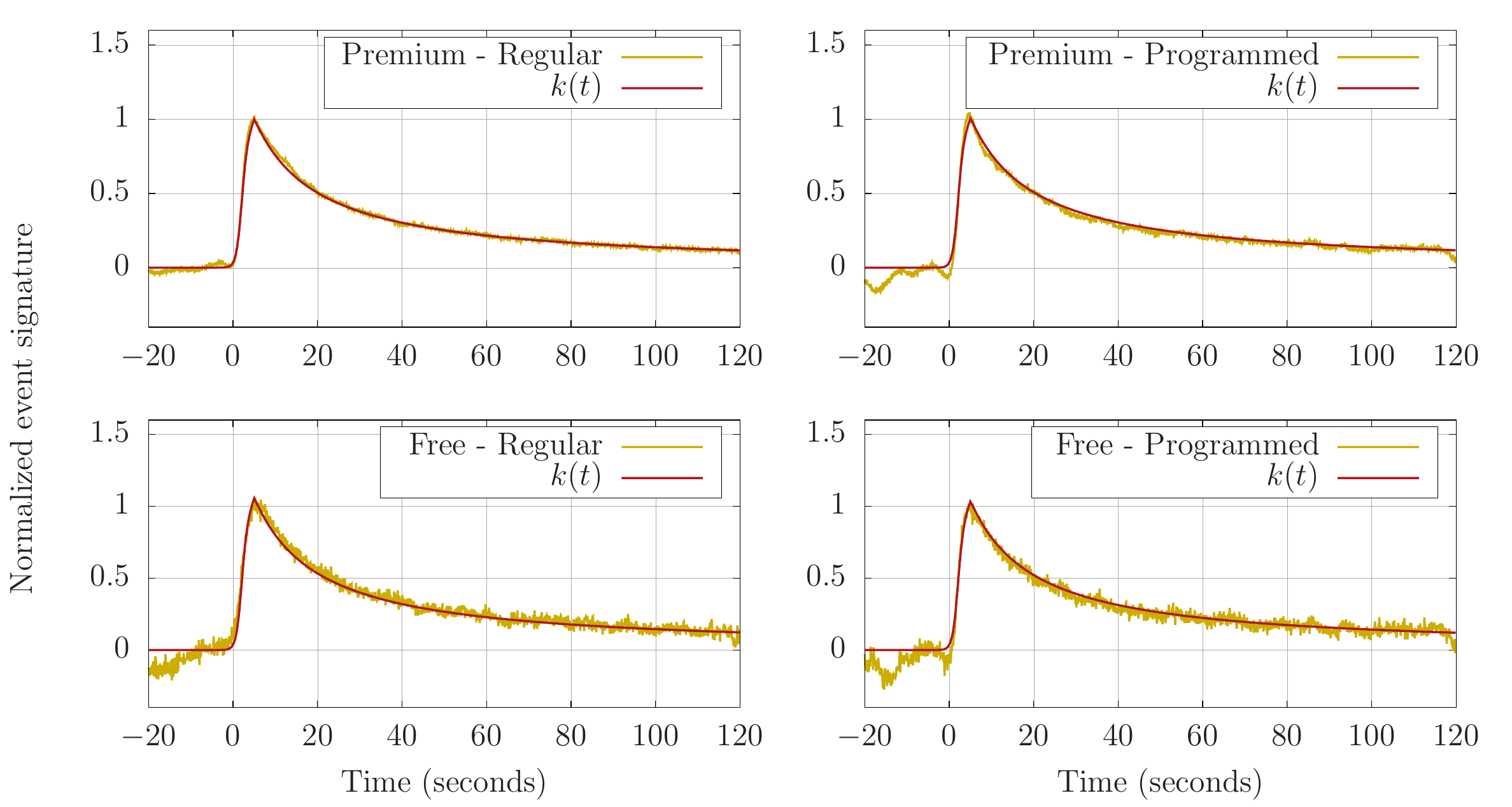}
\caption{\label{fig:kernel_dissection_contexts} Average event signature for different listening contexts. 
}
\end{figure}

We proceed to an analysis of the empirical event signatures across genres and listening contexts, to confirm whether the shape chosen for the event kernels in Eq. (\ref{eq:lambda_explicit}) remains consistent across a wide number of conditions. Figure \ref{fig:kernel_dissection_genres} shows the analysis across a number of genres (R\&B, Rap, Dance \& House, Rock, Indie Rock, Pop). In all cases, the kernel shape closely matches the empirical event signature, showing consistence across genres. Figure \ref{fig:kernel_dissection_contexts} proceeds to the same analysis across listening contexts (free or premium subscription, and whether or not the song was played based on a recommendation by the platform), with similar results.

\clearpage

\section{Cross-sectional analysis of skip profiles}\label{sec:analysis}

We can use the sparse parametrization of skip profile curves to compare skip profiles quantitatively across a number of contexts.

\begin{figure}[!b]
\centering
\begin{subfigure}[b]{\textwidth}
\centering
\includegraphics[width=0.85\textwidth]{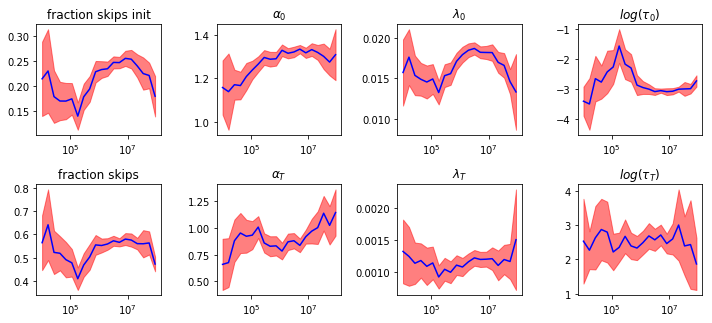}  
\caption{\label{fig:analysis_per_plays} Analysis by number of streams.}
\end{subfigure}
\begin{subfigure}[b]{\textwidth}
\centering
\includegraphics[width=0.85\textwidth]{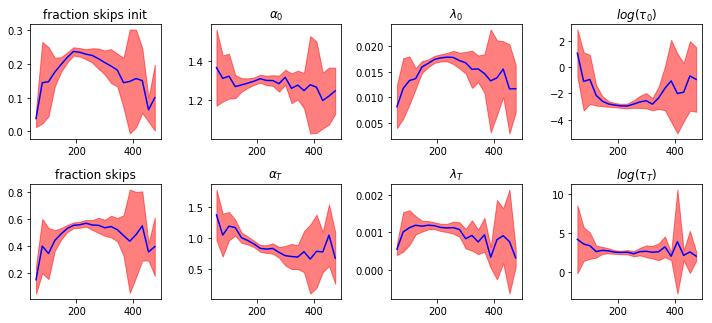}  
\caption{\label{fig:analysis_per_duration} Analysis by song duration.}
\end{subfigure}
\caption{The skips and parameters statistics, as a function of the number of streams and song duration. The blue curve is the mean and the red area represents the $95\%$ confidence interval.}
\end{figure}

\paragraph{Stream count} Figure \ref{fig:analysis_per_plays} shows the parameter distributions as a function of stream count. Most parameters are quite stable, as a horizontal line would almost fit in the $95\%$ confidence interval. A slight trend might be visible for $\alpha_0$ and $\alpha_T$, which tends to increase with the stream count. The parameters relative to the beginning of the track $(\alpha_0, \lambda_0, \tau_0)$ are overall more stable than the parameters relative to the end of the track $(\alpha_T, \lambda_T, \tau_T)$. The intercept $\tau_0$ is consistently very small.

\paragraph{Duration} Figure \ref{fig:analysis_per_duration} shows the parameter distributions as a function of the duration of the song. Here again, parameters are quite stable, except $\alpha_T$ which decreases with increasing durations, showing that the anticipation behaviour changes with the duration of the track -- it is steeper for shorter tracks.

\paragraph{Genres} We finally proceed to an analysis by genres, which confirms the universality of the parameters across genres. Two notable deviations can be observed: (i) classical and jazz songs have a much lower skip rate than other genres, which reflects on the distribution of $\lambda_0$ and $\lambda_T$, (ii) one can guess a bi-modal distribution for $\lambda_T$, with half of the songs being more heavily skipped at the end than the other half.

\begin{figure}[!h]
\centering
\includegraphics[width=\textwidth]{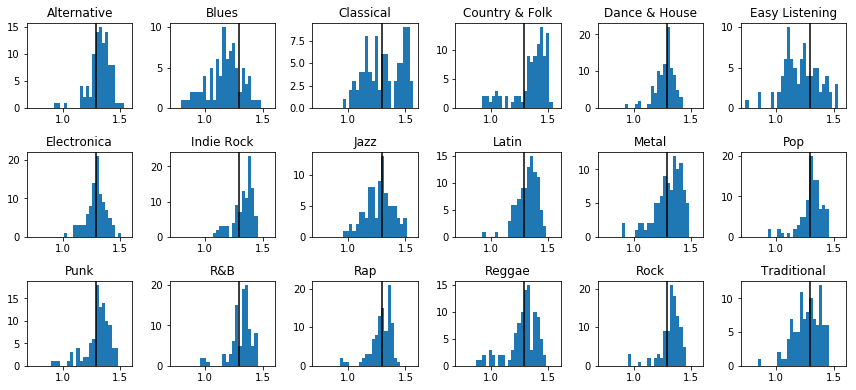}  
\caption{Distribution of $\alpha_0$ per genres.}
\end{figure}

\begin{figure}[!h]
\centering
\includegraphics[width=\textwidth]{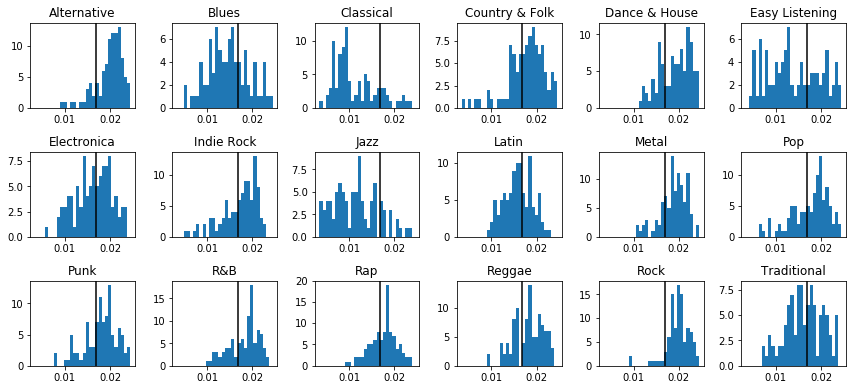}  
\caption{Distribution of $\lambda_0$ per genres.}
\end{figure}

\begin{figure}[!h]
\centering
\includegraphics[width=\textwidth]{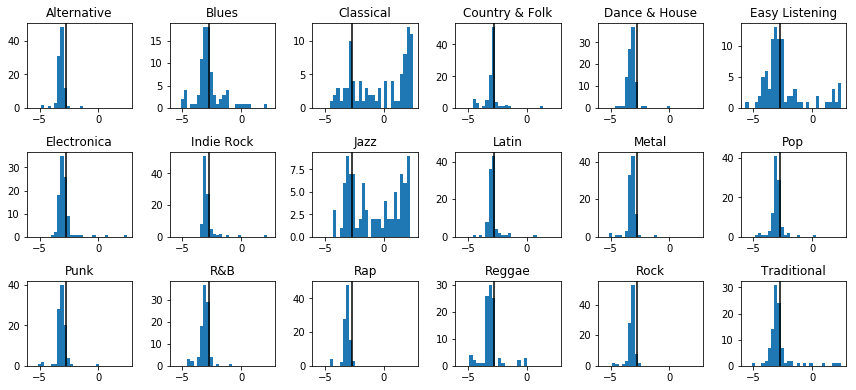}  
\caption{Distribution of $\text{log}(\tau_0)$ per genres.}
\end{figure}

\begin{figure}[!h]
\centering
\includegraphics[width=\textwidth]{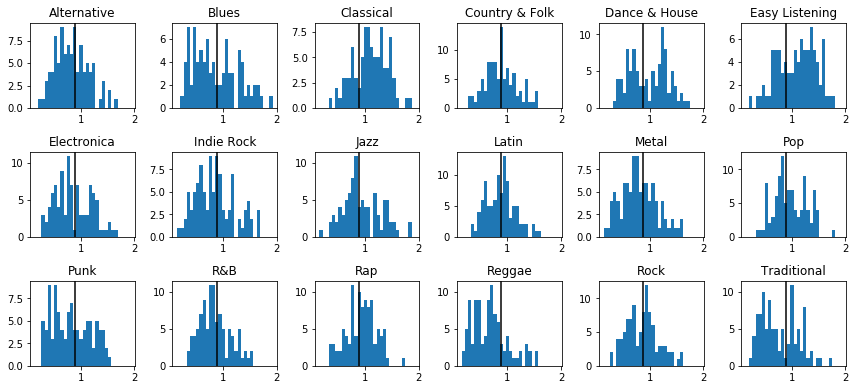}  
\caption{Distribution of $\alpha_T$ per genres.}
\end{figure}

\begin{figure}[!h]
\centering
\includegraphics[width=\textwidth]{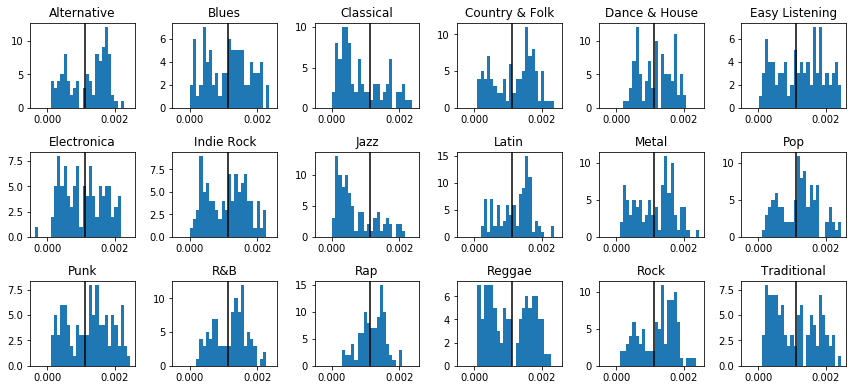}  
\caption{Distribution of $\lambda_T$ per genres.}
\end{figure}

\begin{figure}[!h]
\centering
\includegraphics[width=\textwidth]{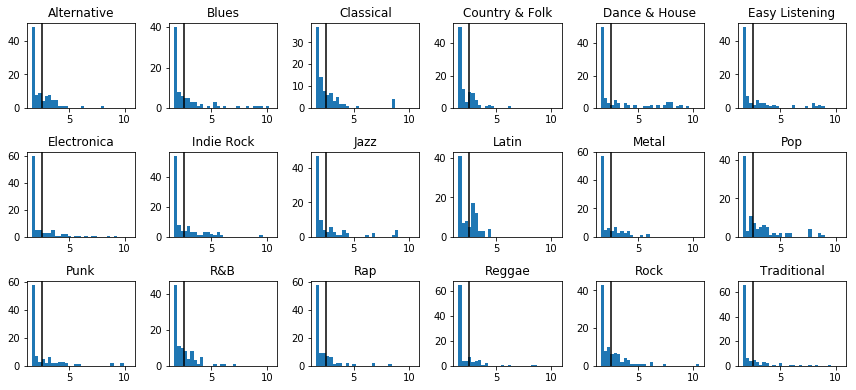}  
\caption{Distribution of $\text{log}(\tau_T)$ per genres.}
\end{figure}



\end{document}